\documentclass[conference,compsoc]{IEEEtran}
\makeatletter
\def\@IEEEauthorrefmark#1{\textsuperscript{#1}}
\makeatother
\ifCLASSOPTIONcompsoc
  \usepackage[nocompress]{cite}
\else
  \usepackage{cite}
\fi

\usepackage{verbatim} 
\usepackage{graphicx}
\usepackage{enumitem}
\usepackage{colortbl} 
\usepackage{xspace}
\usepackage{amsmath}
\usepackage{url}
\usepackage[framemethod=TikZ]{mdframed}
\usepackage[compatibility=false]{caption}  
\usepackage{subcaption}
\usepackage{float}
\usepackage{balance}
\usepackage{multirow} 
\usepackage{booktabs} 
\usepackage{xcolor}  
\usepackage{threeparttable}
\newcommand{\name}{\texttt{CompLeak}\xspace}
\newcommand{\nameappend}{\texttt{CompLeak}}
\newcommand{\mypara}[1]{\smallskip\noindent\textbf{#1}}

\ifCLASSINFOpdf
\else
\fi

\begin{document}
%
\title{\name: Deep Learning Model Compression Exacerbates Privacy Leakage}


\author{
\IEEEauthorblockN{
Na Li\textsuperscript{1},
Yansong Gao\textsuperscript{2},
Hongsheng Hu\textsuperscript{3},
Boyu Kuang\textsuperscript{1}, 
Anmin Fu\textsuperscript{1}
}

\IEEEauthorblockA{\textsuperscript{1}Nanjing University of Science and Technology, China, \{li\_na, kuang, fuam\}@njust.edu.cn}

\IEEEauthorblockA{\textsuperscript{2}The University of Western Australia, Australia, garrison.gao@uwa.edu.au}

\IEEEauthorblockA{\textsuperscript{3}The University of Newcastle, Australia, Hongsheng.Hu@newcastle.edu.au}
}


\maketitle

\begin{abstract}
Model compression is crucial for minimizing memory storage and accelerating inference in deep learning (DL) models, including recent foundation models like large language models (LLMs). Users can access different compressed model versions according to their resources and budget.  
However, while existing compression operations primarily focus on optimizing the trade-off between resource efficiency and model performance, the privacy risks introduced by compression remain overlooked and insufficiently understood.

In this work, through the lens of membership inference attack (MIA), we propose \name, the first privacy risk evaluation framework examining three widely used compression configurations that are pruning, quantization, and weight clustering supported by the commercial model compression framework of Google's TensorFlow-Lite (TF-Lite) and Facebook's PyTorch Mobile. \name has three variants, given available access to the number of compressed models and original model.
\nameappend$_{\rm NR}$ starts by adopting existing MIA methods to attack a single compressed model, and identifies that different compressed models influence members and non-members differently. When the original model and one compressed model are available, \nameappend$_{\rm SR}$ leverages the compressed model as a reference to the original model and uncovers more privacy by combining meta information (e.g., confidence vector) from both models. When multiple compressed models are available with/without accessing the original model, \nameappend$_{\rm MR}$ innovatively exploits privacy leakage info from multiple compressed versions to substantially signify the overall privacy leakage. 
We conduct extensive experiments on seven diverse model architectures (from ResNet to foundation models of BERT and GPT-2), and six image and textual benchmark datasets.
Our experimental results show that \nameappend$_{\rm MR}$ achieves the best MIA performance on all evaluation metrics, including TPR @ 0.1\% FPR, proving that model compression exacerbates privacy leakage.
\end{abstract}


%
\IEEEpeerreviewmaketitle

\section{Introduction}
\label{sec:intro}

Driven by the large-scale data, DL has witnessed remarkable advancements, excelling in applications such as self-driving~\cite{ndikumana2020deep}, protein structure prediction ~\cite{jumper2021highly}, and the recent wave in Artificial Intelligence Generated Content (AIGC), including text generation through ChatGPT~\cite{stokel2023chatgpt} and image generation via diffusion model~\cite{yang2023diffusion}.
However, these outstanding state-of-the-art models are scaled with an increased number of parameters, which require considerable computational resources and memory footprint, challenging their deployment on resource-constrained devices.

Model compression~\cite{cheng2018compression,han2015deep,choudhary2020comprehensive,he2022sparse, hoefler2021sparsity, molchanov2017variational,zhu2024survey} has been a mainstream technique to resolve such challenges, which are already widely used by industries. 
Commercially available compression frameworks, such as Google’s TF-Lite and Facebook’s PyTorch Mobile, facilitate the deployment of DL models on Internet of Things (IoT) and mobile devices~\cite{ma2023quantization}. These frameworks enable model providers to easily generate compressed models using operations like weight clustering, pruning, and quantization, either during training or post-training. Additionally, various toolkits support the compression of foundation models (e.g., LLMs) through methods such as quantization or distillation, helping to mitigate their substantial computational and storage demands~\cite{zhu2024survey,lan2019albert,Zhang2024LoRAPrune,zafrir2019q8bert,zhao2024atom}.
Furthermore, model providers can also employ compression operations to provide interfaces for models with varying model sizes---larger models generally deliver superior performance but are more expensive---enabling users to selectively access customized models according to their needs and budget.

\subsection{Limitation}
While model compression greatly accelerates inference and reduces memory storage, it has been delicately studied and shown to be vulnerable to security attacks such as backdoor attacks~\cite{ma2023quantization,egashira2024exploiting}. However, the {\it privacy risks} imposed by {\it model compression} are overlooked and poorly understood, although the privacy risks of DL models have been widely studied~\cite{li2022auditing,baluta2022membership,ye2022enhanced,li2021membership,shokri2017membership,carlini2022membership,li2024seqmia,he2024difficulty}.

DL models inherently memorize sensitive information from their training datasets, with MIA emerging as a commonly used auditing technique for evaluating such privacy risks~\cite{li2021membership,chen2020gan,he2024difficulty,li2024seqmia}. MIA exploits a model’s tendency to overfit its training data, leading to significant differences in outputs between training set members and non-members. This vulnerability enables attackers to infer whether a given data sample was part of the training dataset, posing a significant threat to individual privacy. For instance, an attacker could deduce that a person participated in a confidential clinical trial by determining that their medical records were used to train a predictive model for an experimental drug. On the other hand, MIA can serve as a valuable tool for auditing privacy leakage, particularly in light of stringent privacy regulations such as the General Data Protection Regulation (GDPR)~\cite{gdpr}, which mandates strong protections for user data.

Notably, model compression operations for DL models have traditionally been employed to balance model capacity and performance, while the privacy risks stemming from compression remain unexplored---especially in scenarios where multiple compressed models are accessible. To our knowledge, the most relevant study is by Li \textit{et al.}~\cite{yuan2022membership}, which assesses the privacy leakage of a pruned model via MIA. However, their evaluation is fundamentally different from our work, which is limited to reliance on information from a single-pruned model and does not account for the unique privacy risks introduced by model compression, where new insights could be derived by correlating information across different compressed model versions and the original model.

To this end, we ask the following research questions to underscore the urgent need for a comprehensive investigation into the privacy risks of mainstream compression technologies.

\begin{mdframed}[backgroundcolor=black!10,rightline=false,leftline=false,topline=false,bottomline=false,roundcorner=2mm]
   Does model compression exacerbate privacy leakage with increasingly access to multiple compressed models? If so, to what extent does it amplify privacy risks?
\end{mdframed}

\subsection{Our Work}
\begin{figure}[t]
    \centering
    \includegraphics[trim=0 0 0 0,clip,width=0.35\textwidth]{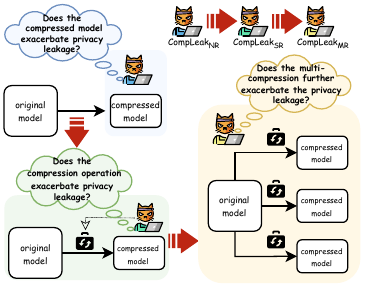} 
    \caption{Compression scenarios targeted by the three \name variants: blue/green/yellow attacker icon stands for \nameappend$_{\rm NR}$/\nameappend$_{\rm SR}$/\nameappend$_{\rm MR}$ considering number of accessible compressed model versions.}
    \label{fig:leak}
\end{figure}

This work, for the first time, unveils and confirms that model compression exacerbates privacy leakage through the lens of MIA as a privacy auditing approach.
The primary reason is that differing compression operations affect members and non-members differently, where different compressed model versions leak privacy in slightly different ways due to variations in their memorization capacity and the inherent randomness of compression operations (e.g., pruning at different or even identical rates, weight clustering with varying or identical numbers of clusters).
Consequently, aggregating leakage from multiple sources amplifies the overall privacy risk. 
To quantify the extent of this additional leakage, we design various MIA methods tailored for a wide range of compression scenarios, as illustrated in Figure~\ref{fig:leak}, primarily considering the number of accessible compressed model versions. \nameappend$_{\rm NR}$, which directly adopts existing MIA techniques, evaluates privacy leakage per compressed model without relying on any reference or paired model. \nameappend$_{\rm SR}$ introduces new MIA techniques to capture additional privacy leakage from a compressed model version paired with the original model. Furthermore, \nameappend$_{\rm MR}$ enhances MIA techniques to assess privacy risks when multiple compressed model versions are accessible, with or without the availability of the original model.
Below, we highlight the key findings of each variant under the \name framework and brief its core attack design.

\mypara{\nameappend$_{\rm NR}$.} 
In the context of \name, we refer to existing MIA~\cite{nasr2018machine,shokri2017membership,yuan2022membership,song2021systematic,yeom2018privacy} (training-based~\cite{nasr2018machine,shokri2017membership,yuan2022membership}, metric-based~\cite{song2021systematic,yeom2018privacy}), which conduct solely based on the leaked information from an underlying model itself---where no reference model is used---as \nameappend$_{\rm NR}$, to audit the vulnerabilities per compressed model with varying compression degrees.
Each compressed model obtained through pruning, quantization, and weight clustering compression operations is evaluated in our experiments upon \nameappend$_{\rm NR}$.
Notably, consistent with the results in~\cite{yuan2022membership}, we observe that highly compressed models are generally less vulnerable than the original model.
For instance, when an MLP-based attack meta-classifier~\cite{nasr2018machine} targets a 90\%-pruned VGG16 model on Mini-ImageNet, the attack accuracy drops by 5\% compared to the original model.
This reduction is potentially because high-level compression significantly limits model capacity and suppresses overfitting~\cite{he2022sparse, hoefler2021sparsity, molchanov2017variational,Wang2021prune}.
On the contrary, pruning with lower sparsity, quantization into 8-bit integers, and weight clustering with more centroids exhibit comparable privacy leakage to the uncompressed model. 

Despite the overall MIA accuracy remaining relatively consistent across different compressed models, the way each compressed model affects members and non-members varies. This variation serves as the foundation for additional privacy leakage, where leaked information is newly captured through \nameappend$_{\rm SR}$ and \nameappend$_{\rm MR}$.

\begin{figure}[t]
    \centering
    \includegraphics[trim=0 0 0 0,clip,width=0.4\textwidth]{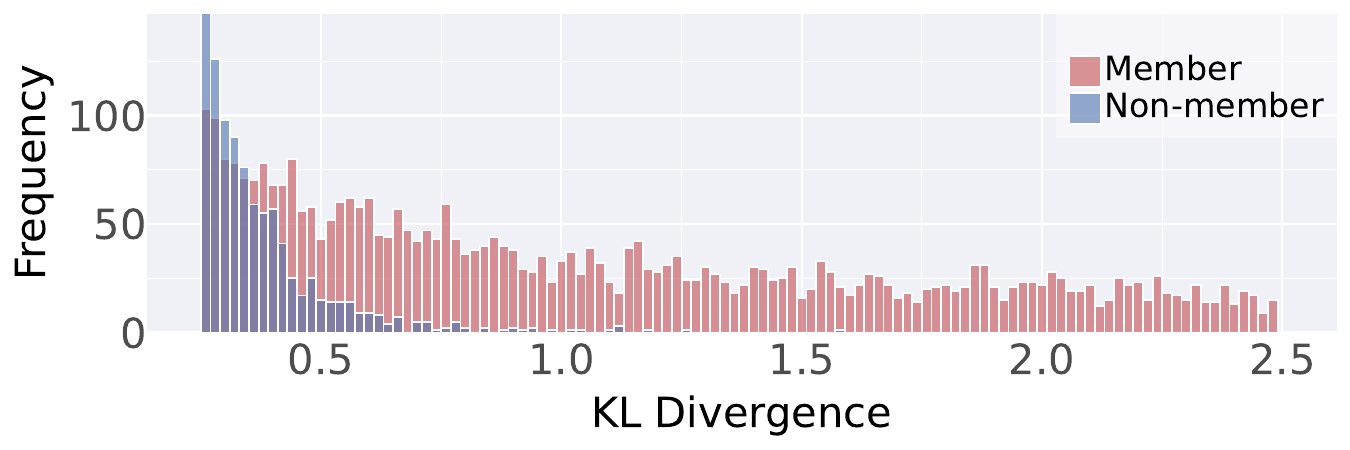}
    \caption{The KL divergence between the two posteriors on the same samples, obtained from the original MobileNetV2 (trained on Tiny-ImageNet) and the 40\% pruned MobileNetV2.}
    \label{fig:kl}
\end{figure}

\mypara{\nameappend$_{\rm SR}$.}
We note that the impact induced by compression operations (e.g., on the posterior probability distribution) varies substantially between members and non-members, as shown in Figure~\ref{fig:kl}. Therefore, our insight is that capturing the subtle alterations imposed by compression operations will amplify privacy.
Based on this intuition, instead of using information from a single model only, we incorporate leaked information from a compressed model and pair it with the information from the original model to improve the MIA performance. We refer to this \name variant as \nameappend$_{\rm SR}$ as a single reference model is utilized.

Generally, \nameappend$_{\rm SR}$ utilizes \textit{meta-data construction} to combine a pair of posteriors, one from the original model and the other from a corresponding compressed version, to form meta-data that is used to train a binary meta-classifier for membership inference.
From the experimental results, regardless of the compression operation's type and degree, \nameappend$_{\rm SR}$ exhibits strong capabilities, providing evidence that model compression indeed threatens privacy. Specifically, \nameappend$_{\rm NR}$~\cite{nasr2018machine} achieves 60\% MIA accuracy on the original VGG16 trained on Mini-ImageNet, which drops to 55\% on the pruned model with 90\% parameter removal. As a comparison, when exploiting the two models together, our \nameappend$_{\rm SR}$ significantly improves the accuracy to 74\%, a 19\% accuracy gain.

\mypara{\nameappend$_{\rm MR}$.}
Model service providers typically release a set of compressed models (i.e., more than one)  with varying capacities via different interfaces~\cite{touvron2023llama,gpt}, so we devise \nameappend$_{\rm MR}$ to exploit multiple compressed models as references to further amplify the privacy leakage.
When we intuitively adopt the same methodology as \nameappend$_{\rm SR}$ to combine the posteriors generated by multiple compressed models for the same target sample as meta-data, no improvement is observed in the attack performance compared to \nameappend$_{\rm SR}$.
We argue that this occurs because \nameappend$_{\rm SR}$ utilizes the variation in posteriors between the original model and a compressed model, whereas the variation in posteriors across multiple versions is minimal and more challenging to interpret.
 
Encouragingly, we discover two remarkable phenomena that support our intuition that different compressed models leak privacy in slightly different ways. 
First, building on \nameappend$_{\rm SR}$, although it becomes complicated to directly capture subtle varieties among the posteriors generated by multiple compressed models, we observe that membership inference results from \nameappend$_{\rm SR}$ attack meta-models---each trained to target a certain level compressed model---on the same target sample exhibit notable differences. 
For example,  the membership status predicted by \nameappend$_{\rm SR}$ shows approximately 22\% discrepancy when targeting an 80\%-pruned VGG16 versus a 90\%-pruned VGG16 on the Mini-ImageNet.
Second, we identify that as the compression degree increases, the evolution of loss calculated from compressed models using ground truth labels and the cross-entropy reveals disparities between members and non-members, both in direction and magnitude. 
Specifically, the loss for members increases with higher sparsity levels, while the loss for non-members fluctuates. 

The above new findings provide us with new insights into the design of \nameappend$_{\rm MR}$, aggregating leaks from multiple sources to further amplify the overall privacy risk.
More concretely, an adversary can first utilize \textit{posterior concatenation} by querying each \nameappend$_{\rm SR}$ attack meta-classifier for the target sample to obtain a set of \nameappend$_{\rm SR}$ attack meta-posteriors, which are then concatenated.
Then, in the \textit{loss concatenation} step, the target sample is fed into each compressed model in ascending order of compression degree to compute a set of losses, which are also concatenated. 
Finally, the concatenated posteriors and losses are stacked to form meta-data, which is used to train a \nameappend$_{\rm MR}$ meta-classifier for membership inference.
Extensive experiments show that multiple compressed models further exacerbate membership leakage compared to \nameappend$_{\rm SR}$ using a single compressed model, especially in TPR @ 0.1\% FPR.
Additionally, we relax the adversary's knowledge, following~\cite{yuan2022membership}, by assuming they can only access multiple compressed versions, but not the original model. In this setting, the adversary cannot obtain \nameappend$_{\rm SR}$ attack meta-posteriors, instead, posterior concatenation refers to concatenating the posteriors obtained by querying each compressed model for the target sample. Experimental results indicate that although \nameappend$_{\rm MR}$ exhibits a decline in performance under this setting, it still outperforms the best \nameappend$_{\rm NR}$ targeting a single compressed or original model.

\mypara{Contribution.} Our main contributions can be summarized as:
\begin{itemize}[leftmargin=*]
    \item We propose \name, the first systematic privacy risk evaluation framework that examines three widely used compression operations---pruning, quantization, and weight clustering---through the lens of membership inference attacks.
    \item We employ the existing MIA as \nameappend$_{\rm NR}$, relying solely on information from the underlying model, to comprehensively assess the privacy leakage per compressed model ({\bf Section~\ref{sec:NR}}).
    \item We present \nameappend$_{\rm SR}$ for a single compression scenario, unveiling that regardless of the compression degree, the compression operations indeed jeopardize privacy ({\bf Section~\ref{sec:SR}}).
    \item We propose \nameappend$_{\rm MR}$ aggregating leaks from multiple compressed models to further amplify the privacy leakage caused by model compression ({\bf Section~\ref{sec:MR}}). 
    \item We conduct extensive experiments in both the classic image domains and the emerging field of foundation models to demonstrate the effectiveness of \name.
\end{itemize}

\mypara{Ethic and Privacy Considerations.}
All our experiments are conducted on publicly available datasets that are widely used in related privacy leakage research, and we strictly adhere to their respective usage licenses.

Although we use commercial toolkits like TensorFlow-Lite for model compression, the observed privacy leakage arises from general model compression techniques, not from any specific tool implementation. This aligns with the findings in~\cite{ma2023quantization}, where TensorFlow Lite was used to demonstrate a backdoor attack by exploiting model compression. The TensorFlow Lite security team acknowledged~\cite{ma2023quantization} that this vulnerability could not be mitigated through changes to the implementation, as it stems from the fundamental design of post-training quantization.

To mitigate such risks, our findings strongly suggest that model providers adopt a range of strategies---such as incorporating differential privacy, training with synthetic data, and reducing model overfitting---before releasing models through query APIs. These practices help safeguard user data and ensure ethical standards in the deployment of model compression techniques.
\section{Background and Related Work}
\label{sec:related}

\subsection{Compression}
\begin{figure}[t]
    \centering
    \includegraphics[trim=0 0 0 0,clip,width=0.40\textwidth]{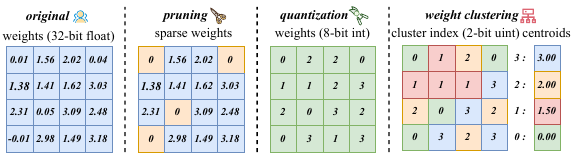} 
    \caption{The same weight matrix subjected to pruning, quantization and weight clustering separately.}
    \label{fig:compress}
\end{figure}

Among the various compression operations, the three that are currently most widely employed and supported by commercial compression frameworks~\cite{han2015deep,jacob2018quantization}, namely \textit{pruning}~\cite{anwar2017structured,he2022sparse, hoefler2021sparsity, molchanov2017variational}, \textit{quantization}~\cite{gholami2022survey,yang2019quantization}, and \textit{weight clustering}~\cite{han2015deep}.
Upon each operation, we illustrate the weight matrix results in Figure~\ref{fig:compress}.

\mypara{Pruning.}
Pruning generally involves removing relatively unimportant parameters, typically following the ``train-prune-finetuning'' workflow~\cite{yuan2022membership}. Currently, it is primarily categorized into unstructured and structured pruning, with the former delivering higher compression rates and prediction accuracy, while the latter offers superior hardware acceleration. Unstructured pruning ignores the model's architecture and focuses on removing individual parameters. For instance, Han \textit{et al.}~\cite{han2015deep} set the parameters with the lowest magnitudes to zero. 
In contrast, structured pruning leverages the model's structure to remove parameters in a more organized manner, typically by removing entire groups of parameters. For example, Li \textit{et al.}~\cite{Li2016PruningFF} removed entire filters with the lowest absolute values from the convolution layers.
Men \textit{et al.}~\cite{men2024shortgpt} observed certain layers that contribute little to the overall performance. Building on this, they used Block Influence (BI) to measure the importance of each layer based on the similarity between its input and output and performed layer pruning by removing layers with low BI values.

\mypara{Quantization.}
Quantization typically converts 32-bit floating-point weight formats to the most compact and popular 8-bit integers, broadly divided into training-aware quantization (QAT) and post-training quantization. The former maintains the model's performance by learning the quantized parameters, but it follows a time-consuming ``train-QAT-finetuning'' workflow~\cite{jacob2018quantization}. The latter avoids the training process by using a small calibration dataset to guide the quantization, allowing it to be directly applied to pre-trained full-precision models~\cite{hubara2021accurate}.
In addition, dynamic quantization belongs to the category of post-training quantization and provides support for LLMs. The system automatically selects the scale factor for activations based on the data range observed at runtime, eliminating the need for additional fine-tuning.

\mypara{Weight Clustering.}
It is also known as weight sharing, which groups the weights of each layer into multiple clusters based on their similarity and assigns the centroid value of each cluster to all the weights within it. In this case, each group only needs to store the centroid value and the corresponding cluster index, as illustrated in Figure~\ref{fig:compress}. During weight updates, gradients for each weight are computed and aggregated within the same cluster to perform the update~\cite{han2015deep}.
For LLMs, Lan \textit{et al.}~\cite{lan2019albert} achieved cross-layer parameter sharing with acceptable model performance degradation.

Note that these three compression operations are directly supported by commercial toolkits, i.e., Google's TF-Lite, where pruning and quantization are also supported by Facebook's PyTorch Mobile, Microsoft's NNI, and NVIDIA's TensorRT. 
In this work, our experiments are mainly implemented on the DL framework PyTorch 2.0.1, with tools including Microsoft's NNI for $L_1$ unstructured pruning, Facebook's PyTorch Mobile for dynamic quantization and QAT, and the PyTorch implementation based on work~\cite{han2015deep} for weight clustering. In addition, we also performed weight clustering using TF-Lite under TensorFlow 2.7.0, as detailed in Appendix~\ref{appendix:cluster}.

\subsection{Membership Inference Attack}

MIA aims to determine whether a target sample is part of the training dataset for a given  model~\cite{li2021membership,li2024analyzing,yuan2022membership,carlini2022membership,li2024seqmia,he2024difficulty}. 
Because of the simplicity of the definition, MIA has been widely used for evaluating the privacy risks of DL models~\cite{li2022auditing,baluta2022membership,ye2022enhanced,tang2022mitigating,shang2025defend}.
Formally, considering a target sample $x$, a trained victim model $\mathcal{M}$, the process of membership inference can be defined as:
\begin{equation}
\mathcal{A}: x, \mathcal{M} \rightarrow \{ 0, 1 \}
\end{equation}
The attack meta-model $\mathcal{A}$ is essentially a binary classifier and can be constructed in various ways. If a target sample $x$ has been used to train $\mathcal{M}$, $\mathcal{A}$ outputs 1 (i.e., member) and 0 otherwise (i.e., non-member).

In practical scenarios, most MIAs are conducted in black-box settings, where adversaries have access only to the posterior probability distribution of the victim model's output. A common strategy proposed by Shokri \textit{et al.}~\cite{shokri2017membership} is to train shadow models that mimic the behavior of the victim model. These shadow models output posterior as meta-data to train a binary meta-classifier for membership inference. To improve attack performance, Nasr \textit{et al.}~\cite{nasr2018machine} incorporated the true labels of data samples as features into the meta-data. 
Additionally, several studies~\cite{song2021systematic,yeom2018privacy,salem2018ml} introduced metric-based MIAs that use metric values, e.g. posterior, entropy, or modified entropy, calculated from the victim model's output to distinguish membership status without the need to train the attack meta-classifier.
Furthermore, some works utilize MIA to investigate the privacy impact of specific technologies or systems, e.g., explainable machine learning methods~\cite{liu2024please}, synthetic data~\cite{zhao2025does}, machine unlearning~\cite{chen2021machine}, visual self-supervised encoders~\cite{zhu2024unified}, speaker recognition systems~\cite{chen2024slmia}, query-based systems~\cite{stevanoski2024querycheetah} and multi-exit networks~\cite{li2022auditing}. 
Recently, MIA has been extended to the foundation models, e.g., diffusion models~\cite{Peng2025diffence,pang2025black} and LLMs~\cite{meeus2024did,wen2024membership}. 

To the best of our knowledge, the work from Yuan \textit{et al.}~\cite{yuan2022membership} is the only one that targets pruning in the context of a compressed model, namely SAMIA. 
However, SAMIA fundamentally differs from our work, as it targets only a single-pruned model. Because it overlooks the unique privacy risks of accessing multiple compressed model versions rooted in model compression, making it essentially the same as other works attacking a full-precision model.

\section{Threat Model}
\label{threat}
This section defines the threat model by providing a detailed description of the adversary's knowledge, capability, and objective.
Notably, the three \name variants, which will be discussed in the following three sections, are all conducted under this threat model.

\mypara{Adversary Knowledge.}
As mentioned earlier, service providers generally release interfaces to models in various sizes allowing users to selectively access.
In this paper, we focus on the most commonly adopted black-box setting as in existing works~\cite{pang2025black,chen2024slmia,li2024seqmia,he2024difficulty,meeus2024did}, where the attacker only has access to the posterior probability distribution of the victim's outputs from the original model and its associated compressed models.
Moreover, we relax this assumption in Section~\ref{threat_MR}, where the original model is inaccessible.
We also align with SOTAs~\cite{li2024seqmia,he2024difficulty,yuan2022membership}, assuming that the adversary has a local shadow dataset $\mathcal{D}_{s}$, which has the same distribution as the victim’s training dataset but without any overlap.
The adversary also knows the architecture of the victim (original/compressed) model---accurate model architecture can be stolen in a black-box manner via side-channel information~\cite{yan2020cache,gao2024deeptheft}, as well as the compression configuration---this information is often provided by the model provider, and the ground truth of the target samples whose membership status needs to be inferred.

\mypara{Adversary Capability.}
The adversary can utilize $\mathcal{D}_{s}$ to train a set of shadow (original/compressed) models that share the same architecture as the victim (original/compressed) models, to mimic the victim's behavior.
 
\mypara{Adversary Goal.}
Given a target sample, the adversary aims to determine whether it belongs to the training dataset of the victim model according to the above knowledge and ability.

\section{\nameappend$_{\rm NR}$}
\label{sec:NR}

In this section, we directly utilize existing membership inference attacks~\cite{shokri2017membership,nasr2018machine,song2021systematic,yeom2018privacy,yuan2022membership}, which are based solely on leaked information from the target model---without using a reference model---denoted as \nameappend$_{\rm NR}$, to quantify and compare the privacy risks between the original model and each compressed model with varying compression degrees and configurations.
Since the only difference among the MIA adopted in \nameappend$_{\rm NR}$ is how they leverage the underlying model's leaked information, we begin by categorizing and describing their attack methodologies. Then, we present the evaluation results.
It is crucial to emphasize that our goal here is to assess the privacy leakage of a given compressed model through \nameappend$_{\rm NR}$, rather than designing the novel MIA specific to compression scenarios, which will be advanced in Section~\ref{sec:SR} and Section~\ref{sec:MR}.

\subsection{Attack Methodology}
It is important to emphasize that, although the threat model assumes the adversary can access models of varying sizes simultaneously, \nameappend$_{\rm NR}$ only uses the information from a single (whether compressed or not) model to quantify the extent of its membership leakage. It serves as a baseline to understand how compression exacerbates privacy leakage per compressed model being available. 
Here, the \nameappend$_{\rm NR}$ we employ involves five different MIAs, three of them are training-based, and two are metric-based.

\mypara{Training-based.}
In general, training-based attacks require the adversary to train shadow models and leverage the information from these shadow models to construct meta-data. Based on this, the adversary can then train a meta-classifier for membership inference.
Various approaches exist for constructing meta-data, as detailed below: Shokri \textit{et al.} \cite{shokri2017membership} utilize posteriors as meta-data, whereas Nasr \textit{et al.} \cite{nasr2018machine} concatenate ground-truth labels and posteriors for the same purpose. Furthermore, Yuan \textit{et al.} \cite{yuan2022membership} design a SAMIA specifically for a single-pruned neural network, combining the posterior, sensitivity, and ground-truth label as meta-data to train a transformer-based meta-classifier.

\mypara{Metric-based.}
Metric-based attacks use a threshold to infer membership based on the metric values calculated from the victim model's output, without the need for training the attack meta-classifier. In this work, we consider two metrics: entropy loss~\cite{yeom2018privacy} and modified entropy \cite{song2021systematic}.

\subsection{Experiment Setup}

\begin{table}[!t]
\centering
\caption{Classification accuracy under original and three compression operations across different datasets and model architectures.}
\label{acc_result}
\scalebox{0.6}{
\begin{threeparttable}
\begin{tabular}{c|cccccccccc}
\toprule
\multirow{2}{*}{Dataset} & \multicolumn{2}{c}{Original} & \multicolumn{4}{c}{Pruning} & \multicolumn{1}{c}{Quantization} & \multicolumn{3}{c}{Clustering} \\
\cmidrule(lr){2-3} \cmidrule(lr){4-7} \cmidrule(lr){8-8} \cmidrule(lr){9-11}
& Train & Test & 60\% & 70\% & 80\% & 90\% & int-8 & 16 & 8 & 4 \\ 
\midrule
CIFAR-10 & 99.9\% & 70.4\% & 71.6\% & 71.3\% & 71.0\% & 68.6\% & 70.4\% & 71.1\% & 70.0\% & 67.9\% \\
CIFAR-100 & 100\% & 69.3\% & 69.3\% & 69.4\% & 69.1\% & 68.5\% & 69.3\% & 69.2\% & 68.4\% & 66.2\% \\
Mini-ImageNet & 91.7\% & 74.1\% & 73.8\% & 73.7\% & 73.6\% & 73.2\% & 74.1\% & 73.7\% &72.7\% &70.8\%\\
Tiny-ImageNet & 78.9\% & 53.9\% & 53.1\% &53.0\% & 53.0\% & 52.9\% & 53.5\% & 52.8\% & 51.3\% & 44.2\% \\
Location & 100\% & 60.9\% &  60.5\% & 59.7\% & 59.1\% & 56.3\% & 61.1\% & 60.0\% & 58.2\% & 55.8\% \\
\bottomrule
\end{tabular}
\begin{tablenotes}
\small
\item Model architectures: CIFAR-10 (ResNet18), CIFAR-100 (ResNet50), Mini-ImageNet (VGG16), Tiny-ImageNet (MobileNetV2), Location (FCN).
\item For Tiny-ImageNet, the pruning levels are set to $L = \{40\%, 50\%, 60\%, 70\%\}$, as higher ratio degraded model usability.
\end{tablenotes}
\end{threeparttable}}
\end{table}

\begin{table*}[!t]
\centering
\caption{Attack performance of different attacks on varying pruned rate (VGG16+Mini-ImageNet).}
\label{pruning_result}
\scalebox{0.65}{
\begin{threeparttable}
\begin{tabular}{c|ccccccccccccccc}
\toprule
Attack  & \multicolumn{5}{c}{TPR @ 0.1\% FPR (\%)} & \multicolumn{5}{c}{Balanced Accuracy (\%)}  & \multicolumn{5}{c}{AUC (\%)} \\
\cmidrule(r){2-6} \cmidrule(r){7-11} \cmidrule(r){12-16}Method &original & 60\% & 70\% & 80\% & 90\%  &original & 60\% & 70\% & 80\% & 90\%  &original & 60\% & 70\% & 80\% & 90\% \\ 
    
\toprule
    \nameappend$_{\rm NR}$~\cite{shokri2017membership} (LR)  
    &0.0 &0.0 &0.0 &0.0 &0.0
    &48.3 &48.9 &47.1 &48.5 &50.0
    &47.3 &47.9 &45.8 &48.0 &48.0   \\
    \nameappend$_{\rm NR}$~\cite{nasr2018machine} (LR)
    &0.0 &0.0 &0.0 &0.0 &0.0
    &48.3 &50.8 &49.6 &50.0 &51.6
    &50.1 &50.7 &50.2 &51.0 &51.4  \\
    \nameappend$_{\rm NR}$~\cite{shokri2017membership} (RF)   
    &\textbf{1.6} &1.6 &1.5 &1.4 &1.1
    &59.3 &59.4 &59.1 &58.6 &57.8
    &63.2 &63.1 &62.7 &62.1 &60.7 \\
    \nameappend$_{\rm NR}$~\cite{nasr2018machine} (RF)
    &1.5 &1.4 &1.4 &1.4 &1.0
    &59.2 &59.3 &59.2 &58.7 &57.5
    &63.3 &63.3 &62.8 &62.3 &60.5 \\
    \nameappend$_{\rm NR}$~\cite{yeom2018privacy}
    &0.1 &0.0 &0.0 &0.0 &0.0
    &56.7 &57.0 &56.2 &54.2 &52.9
    &54.8 &54.9 &54.2 &52.9 &50.4 \\
    \nameappend$_{\rm NR}$~\cite{song2021systematic}
    &0.1 &0.1 &0.1 &0.1 &0.1
    &59.6&59.8&59.1&58.7&57.3
    &58.7&58.8&58.2&57.1&54.7 \\
    \nameappend$_{\rm NR}$~\cite{yuan2022membership}   
    &0.9 &0.7 &0.7 &0.8 &0.7
    &\textbf{61.9}&61.7&61.4&60.5&59.7
    &\textbf{66.0}&66.4&64.7&64.8 &63.3 \\ 
    \midrule
    \nameappend$_{\rm SR}$ 1 (LR)
    &- &11.3&11.1 &10.6 &8.7
    &- &59.5 &59.7 &59.3 &59.4
    &- &67.1 &67.6 &66.5 &66.8  \\
    \nameappend$_{\rm SR}$ 2 (LR)   
    &- &8.8&8.6 &8.5 &8.0
    &- &59.7 &59.9 &59.9 &60.3
    &- &68.4 &68.5 &68.8 &69.4 \\ 
    \nameappend$_{\rm SR}$ 1 (RF)
    &- &\textbf{39.0}&\textbf{42.1} &\textbf{34.5} &\textbf{25.5}
    &- &\textbf{84.1}&\textbf{84.2} &\textbf{83.4} &79.8
    &- &93.0 &92.8 &91.8 &88.0    \\
    \nameappend$_{\rm SR}$ 2 (RF) 
    &- &36.4 &41.7 &34.2  &24.2
    &- &83.9 &83.9 &83.2 &\textbf{79.9}
    &- &\textbf{93.2}&\textbf{93.0}&\textbf{92.1} &\textbf{88.6} \\ 
\bottomrule
\end{tabular}
\begin{tablenotes} 
\small
    \item \nameappend${\rm SR}$ 1 is based on the first meta-data construction method, while \nameappend${\rm SR}$ 2 is based on the second.
    \item The parentheses represent the structure of the attack meta-model used.
\end{tablenotes} 
\end{threeparttable}}
\end{table*}

\mypara{Datasets.}
Following previous pioneering work~\cite{shokri2017membership,nasr2018machine,yuan2022membership,li2024seqmia,he2024difficulty}, we consider five benchmark datasets covering two data modalities.
Specifically, four image datasets are chosen: CIFAR-10~\cite{CIFAR}, CIFAR-100~\cite{CIFAR}, Mini-ImageNet, and Tiny-ImageNet, while Location, which relates to social connections that contain sensitive personal information in real-world scenarios, represents the text modality~\cite{shokri2017membership}.
A detailed description of all datasets can be found in Appendix~\ref {appendix:datasets}.

\mypara{Victim Model.}
For the image datasets, we utilize four broadly adopted architectures to simulate for the victim (original or compressed) model: ResNet18 \cite{he2016deep}, ResNet50 \cite{he2016deep}, VGG16 \cite{simonyan2014very}, MobileNetV2~\cite{sandler2018mobilenetv2}. For the Location, we train a model with two fully connected layers (FCN), and the implementation details can be found in Appendix~\ref{appendix:fcn}.

\mypara{Meta-classifier.}
Following prior work \cite{chen2021machine}, we adopt four widely employed binary classifiers as the attack meta-classifier: logistic regression (LR), decision tree (DT), multi-layer perceptron (MLP), and random forest (RF), to examine how the meta-classifier with varying capabilities affects MIA performance. 
Due to space limitations, we present only the results for LR and RF, representing the weakest and strongest meta-classifiers.

\mypara{Metric.}
We consider two average-case metrics of balanced accuracy and AUC, along with the TPR @ low FPR proposed by Carlini \textit{et al.}~\cite{carlini2022membership}, all widely used in existing studies~\cite{li2024seqmia,he2024difficulty,liu2022membership,ye2022enhanced}.
\begin{itemize}[leftmargin=*]
    \item \textbf{Balanced Accuracy.} Balanced accuracy measures the probability that an MIA correctly predicts the membership status of samples in a balanced set of members and non-members.
    
    \item \textbf{AUC.} AUC is the area under the receiver operating characteristic (ROC) curve \cite{sankararaman2009genomic} indicates the average success of membership inference.
    
    \item  \textbf{TPR @ low FPR.}  Carlini \textit{et al.}~\cite{carlini2022membership} note that high balanced accuracy/AUC are mainly due to identifying non-members. They recommend to report TPR @ low FPR, which evaluates the true-positive rate at a low false-positive rate (e.g., 0.1\% FPR), providing a more reliable measure of privacy leakage.    
\end{itemize}

\mypara{Original Model Training Settings.}
To mitigate model overfitting, we employed mechanisms including $L_2$ regularization~\cite{krogh1991simple} and early stopping~\cite{prechelt2002early}. We present the training and test accuracy of the original model in Table~\ref{acc_result}. Specifically, for training on CIFAR-10, we follow the experimental setup in~\cite{yuan2022membership}.

\mypara{Model Compression Settings.} Pruning, quantization, and weight clustering are three compression operations widely supported by the commercial framework , which we consider.
\begin{itemize}[leftmargin=*]
    \item \textbf{Pruning.} We apply \textit{L1 unstructured pruning} provided by Microsoft's NNI~\cite{nni} toolkit on the original model at four sparsity levels: $L = \{60\%, 70\%, 80\%, 90\%\}$, which represent the removal of 60\%, 70\%, 80\%, and 90\% of the parameters with the lowest absolute values from the model~\cite{han2015deep}. In addition, we follow the standard pruning workflow, which consists of the stages: ``train-prune-finetune''~\cite{yuan2022membership,anwar2017structured,he2022sparse}. Table~\ref{acc_result} depicts the classification accuracy of the pruned models at these different sparsity degrees. We observe that the pruned versions maintain performance comparable to the original model, with only a minimal drop in accuracy as sparsity increases. Notably, in some cases, the accuracy even shows a slight improvement~\cite{blalock2020state}. 
    \item \textbf{Quantization.} We choose \textit{QAT} supported by Facebook's PyTorch Mobile as the typical quantization operation due to its superior performance in maintaining model accuracy. In practice, converting a float 32-bit original model to an int 8-bit quantized model is the most common setting, as int 4-bit conversion often results in significant performance degradation. For example, both PyTorch Mobile and TF-Lite offer QAT support limited to int 8-bit but not supporting int 4-bit. Therefore, we limit our evaluation to converting an original model to an int 8-bit quantized version. As shown in Table~\ref{acc_result}, the accuracy of the model after QAT remains virtually unchanged.
    \item \textbf{Weight clustering.} We apply weight clustering to the original model's convolutional and linear layers. Specifically, we leverage the K-nearest neighbors (KNN) algorithm to partition the weights of each layer into $N=\{4,8,16\}$ clusters. Table~\ref{acc_result} shows the clustered model's prediction accuracy at different cluster counts, with accuracy dropping as expected as the number of clusters decreases.
\end{itemize}

\subsection{Evaluation Results}
\label{NR_E}

\begin{table}[t]
\centering
\caption{Attack performance of different attacks on original and quantized model (VGG16+Mini-ImageNet).}
\label{qat_result}
\scalebox{0.62}{
\begin{tabular}{c|cccccc}
\toprule
Attack  & \multicolumn{2}{c}{TPR @ 0.1\% FPR (\%)} & \multicolumn{2}{c}{Balanced Accuracy (\%)}  & \multicolumn{2}{c}{AUC (\%)} \\
\cmidrule(r){2-3} \cmidrule(r){4-5} \cmidrule(r){6-7}Method & original &int-8 &original & int-8 & original & int-8 \\     
\toprule
    \nameappend$_{\rm NR}$~\cite{shokri2017membership} (LR)  
    &0.0 &0.0 
    &48.3 &48.3 
    &47.3 &47.4  \\
    \nameappend$_{\rm NR}$~\cite{nasr2018machine} (LR)
    &0.0 &0.0
    &48.3 &51.1 
    &50.1 &50.5   \\
    \nameappend$_{\rm NR}$~\cite{shokri2017membership} (RF)   
    &\textbf{1.6} &1.3
    &59.3 &59.4
    &63.2 &63.4 \\
    \nameappend$_{\rm NR}$~\cite{nasr2018machine} (RF)
    &1.5 &1.2 
    &59.2 &59.3 
    &63.3 &63.5  \\
    \nameappend$_{\rm NR}$~\cite{yeom2018privacy}
    &0.1 &0.1 
    &56.7 &58.2 
    &54.8 &55.3  \\
    \nameappend$_{\rm NR}$~\cite{song2021systematic}
    &0.1 &0.1 
    &59.6 &59.6 
    &58.7 &59.1  \\
    \nameappend$_{\rm NR}$~\cite{yuan2022membership}   
    &0.9 &0.5 
    &\textbf{61.9} &61.1 
    &\textbf{66.0} &64.5  \\
    \midrule
    \nameappend$_{\rm SR}$ 1 (LR)
    &- &23.1 
    &- &60.8
    &- &71.1  \\
    \nameappend$_{\rm SR}$ 2 (LR)   
    &- &10.0 
    &- &59.6 
    &- &70.8  \\
    \nameappend$_{\rm SR}$ 1 (RF)
    &- &\textbf{81.0} 
    &- &\textbf{91.1} 
    &- &\textbf{98.3}  \\
    \nameappend$_{\rm SR}$ 2 (RF) 
    &- &80.7 
    &- &90.3
    &- &\textbf{98.3}  \\
\bottomrule
\end{tabular}}
\end{table}

\begin{table*}[t]
\centering
\caption{Attack performance of different attacks on varying number of clusters (VGG16+Mini-ImageNet).}
\label{basic attack cluster}
\scalebox{0.7}{
\begin{tabular}{c|cccccccccccc}
\toprule
Attack  & \multicolumn{4}{c}{TPR @ 0.1\% FPR (\%)} & \multicolumn{4}{c}{Balanced Accuracy (\%)}  & \multicolumn{4}{c}{AUC (\%)} \\
\cmidrule(r){2-5} \cmidrule(r){6-9} \cmidrule(r){10-13}Method &original & 16 & 8 & 4 &original & 16 & 8 & 4 &original & 16  & 8 & 4 \\ 
    
\toprule
    \nameappend$_{\rm NR}$~\cite{shokri2017membership} (LR)  
    &0.0 &0.0 &0.0 &0.0 
    &48.3 &49.3 &47.4 &51.2 
    &47.3 &47.8 &48.0 &48.7  \\
    \nameappend$_{\rm NR}$~\cite{nasr2018machine} (LR)
    &0.0 &0.0 &0.0 &0.0 
    &48.3 &50.6 &49.7 &50.8 
    &50.1 &50.8 &50.8 &50.4   \\
    \nameappend$_{\rm NR}$~\cite{shokri2017membership} (RF)   
    &\textbf{1.6} &1.4 &1.2 &1.4 
    &59.3 &59.1 &58.5 & 57.9
    &63.2 &63.2 &62.4 & 61.7 \\
    \nameappend$_{\rm NR}$~\cite{nasr2018machine} (RF)
    &1.5 &1.3 &1.2 &1.3 
    &59.2 &58.6 &57.9 &57.4 
    &63.3 &62.8 &62.2 &61.6  \\
    \nameappend$_{\rm NR}$~\cite{yeom2018privacy}
    &0.1 &0.1 &0.1 &0.0 
    &56.7 &56.5 &56.0 &54.4 
    &54.8 &54.8 &54.7 &53.7  \\
    \nameappend$_{\rm NR}$~\cite{song2021systematic}
    &0.1 &0.1 &0.1 &0.1
    &59.6 &60.1 &59.6 &59.3 
    &58.7 &58.9 &58.6 &58.3  \\
    \nameappend$_{\rm NR}$~\cite{yuan2022membership}   
    &0.9 &1.3 &0.6 &0.7 
    &\textbf{61.9} &61.7 &61.5 &61.2 
    &\textbf{66.0} &64.3 &64.1 &62.9 \\ 
    \midrule
    \nameappend$_{\rm SR}$ 1 (LR)
    &- &16.4 &11.3 &8.1
    &- &60.4 &60.1 &59.3
    &- &69.6 &68.3 &65.0  \\
    \nameappend$_{\rm SR}$ 2 (LR)   
    &- &10.0 &8.1 &6.2
    &- &59.9 &59.9 &60.2
    &- &70.8 &68.4 &67.3 \\ 
    \nameappend$_{\rm SR}$ 1 (RF)
    &- &\textbf{67.2} &\textbf{47.3} &\textbf{28.5}
    &- &\textbf{93.2} &89.7 &84.1
    &- &\textbf{98.7} &96.8 &92.4    \\
    \nameappend$_{\rm SR}$ 2 (RF) 
    &- &66.3 &46.8 &\textbf{28.5}
    &- &\textbf{93.2} &\textbf{90.1} &\textbf{84.5}
    &- &\textbf{98.7} &\textbf{97.0} &\textbf{92.8} \\ 
\bottomrule
\end{tabular}
}
\end{table*}

Due to space constraints, unless otherwise specified, the results in the main text are based on the VGG16 trained on Mini-ImageNet. Evaluation results on other datasets can be found in Appendix~\ref{appendix:evaluation}.

\mypara{Pruning Results.}
As shown in Table~\ref{pruning_result}, the performance of \nameappend$_{\rm NR}$ (except for using the low-capability LR as the meta-classifier) on pruned models exhibits a declining trend as the sparsity increases. 
Furthermore, most \nameappend$_{\rm NR}$ achieve comparable performance against pruned models with lower sparsity levels (e.g., 0.6, 0.7) to their performance on the original model,  but their effectiveness reveals a marked decrease when targeting the highly pruned model (e.g., 0.9).
We attribute these to the reduced model capacity, which limits its ability to capture members' details, while the generalization on non-members remains largely unchanged, making the behavior of members more similar to non-members.

\mypara{Quantization Results.}
As detailed in Table~\ref{qat_result}, similar to the attack results on low sparsity pruned models, most \nameappend$_{\rm NR}$ exhibit nearly identical attack performance between the 8-bit quantized model and the original model, with the balanced accuracy difference predominantly below 1\%.

\mypara{Weight Clustering Results.}
Similar to pruning, we observe that the attack performance of \nameappend$_{\rm NR}$ declines with fewer clusters due to the increasing similarity between member and non-member behavior as model capacity reduces. Specifically, when the number of clusters is 4, \nameappend$_{\rm NR}$ shows lower privacy leakage on the clustered model compared to the original one, and when there are 16 clusters, the leakage is nearly identical.

\mypara{Summary.}
These above results indicate that by solely relying on the relationship between members and non-members of the underlying model, without any reference information, the privacy leakage in the compressed model is, in most cases, comparable to that in the original model. Surprisingly, the highly compressed version is less vulnerable to \nameappend$_{\rm NR}$.

\section{\nameappend$_{\rm SR}$}
\label{sec:SR}

After quantifying the membership leakage of the compressed model
through \nameappend$_{\rm NR}$, which utilizes information from only a target compressed version, in this section, we focus on the privacy leakage due to the 
compression operation.
This is achieved through \nameappend$_{\rm SR}$, in which the core idea is to treat the single compressed model as a reference, consistently pairing it with the corresponding original model, to capture the variations caused by the compression operation, i.e., the impact of compression operation on posteriors. Next, we present the design insight, the detailed pipeline of \nameappend$_{\rm SR}$, evaluation results, and the discussion.

\subsection{Design Insight}

We hypothesize that the compression operation affects members and non-members differently, although it is not obvious if we only look at the overall MIA accuracy given a specific compressed model version, as we showed in the \nameappend$_{\rm NR}$.
To prove our hypothesis, we utilize the KL divergence~\cite{kullback1951information} to visualize the distance between the two posteriors on the same target samples, one obtained from the original model and the other from one of its paired compressed models.
As shown in Figure~\ref{fig:kl}, when 40\% of the parameters are pruned from MobileNetV2 on the Tiny-ImageNet, the influence of the compression operation on posteriors (i.e., the changes after compression) is more noticeable for members than non-members.
Essentially, \nameappend$_{\rm NR}$ on the pruned model exhibits a 1.1\% lower balanced accuracy compared to the original model.
This is because, despite fine-tuning, the reduced model capacity, compared to the uncompressed model, is unable to capture the fine-grained features of members. However, the change in generalization is relatively minor, leading to a smaller impact on non-members.

\subsection{Attack Methodology}

\begin{figure}[t]
    \centering
    \includegraphics[trim=0 0 0 0,clip,width=0.46\textwidth]{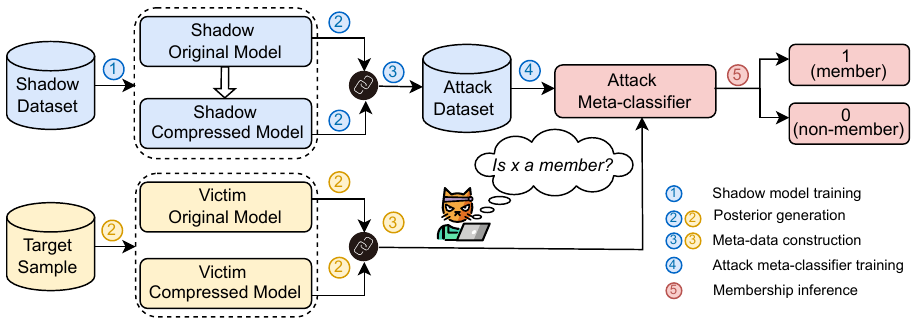}
    \caption{Attack pipeline overview of \nameappend$_{\rm SR}$.}
    \label{fig:workflow}
\end{figure}

Building on the above findings, we propose \nameappend$_{\rm SR}$ specifically for the single-compression scenario, where the key principle is to take a single compressed model as the reference, with the paired model always being the original model.
More specifically, we combine a pair of posteriors from a single compressed model and the original model as meta-data through \textit{meta-data construction} to train the attack meta-classifier for membership inference, capturing the different influences on members and non-members in two aspects: the inherent discrepancy in each posterior, as reflected by \nameappend$_{\rm NR}$~\cite{shokri2017membership,nasr2018machine,yuan2022membership}, and the differences between the two posteriors, which reflect the differential changes caused by the compression operation.

\nameappend$_{\rm SR}$ attack consists of five stages as shown in Figure~\ref{fig:workflow}: shadow model training, posterior generation, meta-data construction, attack meta-classifier training, and attack meta-classifier membership inference. The first five stages are performed once during offline, while the last stage corresponds to the online phase. 

\mypara{Shadow Model Training.}
As mentioned in Section~\ref{threat}, the adversary possesses a shadow dataset $\mathcal{D}_{s}$. The adversary starts by dividing it into two disjoint subsets: the shadow train set $\mathcal{D}_{\rm train}^{s}$, and the shadow test set $\mathcal{D}_{\rm test}^{s}$. The adversary trains a shadow original model $\mathcal{M}_{o}^{s}$ on $\mathcal{D}_{\rm train}^{s}$, then subjects it to a compression algorithm to produce a shadow compressed model $\mathcal{M}_{c}^{s}$ as the shadow reference model. 

\mypara{Posterior Generation.}
The adversary queries both $\mathcal{M}_{o}^{s}$ and $\mathcal{M}_{c}^{s}$ for each data sample from $\mathcal{D}_{s}$, obtaining one pair of posteriors as $\mathcal{P}_{o}^{s}$ and $\mathcal{P}_{c}^{s}$, respectively.

\mypara{Meta-data Construction.}
Inspired by and following ~\cite{chen2021machine}, we use \textit{meta-data construction} to obtain meta-data and provide two construction methods here. In order to better understand posteriors, we first need to sort $\mathcal{P}_{o}^{s}$ in descending order and apply this order to $\mathcal{P}_{c}^{s}$, obtaining $\mathcal{P}_{o}^{s}{'}$ and $\mathcal{P}_{c}^{s}{'}$, respectively~\cite{ganju2018property,chen2021machine}.
\begin{itemize}[leftmargin=*]
    \item The first method is to directly concatenate $\mathcal{P}_{o}^{s}{'}$ and $\mathcal{P}_{c}^{s}{'}$ as meta-data, i.e., $\mathcal{P}_{o}^{s}{'} \parallel \mathcal{P}_{c}^{s}{'}$, where $\parallel$ denotes the concatenation operation.
    \item Since the adversary can access the ground truth label of the audited sample, which has been proven in previous works that the divergence between members and non-members varies in a fine-grained manner across different classes~\cite{yuan2022membership,nasr2018machine}. Thus, the second method is to apply one-hot encoding on the ground truth label to generate $y$, and then concatenate it with $\mathcal{P}_{o}^{s}{'}$ and $\mathcal{P}_{c}^{s}{'}$ as meta-data, i.e., $\mathcal{P}_{o}^{s}{'} \parallel \mathcal{P}_{c}^{s}{'} \parallel y$.
\end{itemize}
 
Note that additional meta-data construction methods (i.e., direct concatenation, $L_{2}$ distance-based) have been considered. Their details and corresponding attack performance can be found in Appendix~\ref{appendix:sc}.

\mypara{Attack Meta-classifier Training.}
The adversary labels the meta-data $X_{a}$ as 1 if it originates from $\mathcal{D}_{\rm train}^{s}$, and as 0 from $\mathcal{D}_{\rm test}^{s}$. 
These labels, denoted as $Y_{a}$, along with $X_{a}$, constitute the attack training dataset used to train the attack meta-classifier $\mathcal{M}_{\rm SR}$, a binary classifier for membership inference. 
During training, binary cross-entropy loss is applied to compute the loss, with the objective of minimizing $\mathcal{L} (\mathcal{M}_{\rm SR}(X_{a}, Y_{a}))$.

\mypara{Attack Meta-classifier Membership Inference.}
Once $\mathcal{M}_{\rm SR}$ is trained, the adversary can determine the membership of a given target sample.
To achieve this, the adversary queries both the victim's original model and a single compressed model to obtain paired posteriors, denoted as $\mathcal{P}_{o}^{v}$ and $\mathcal{P}_{c}^{v}$. 
These paired posteriors are then processed through a meta-data construction step and fed into $\mathcal{M}_{\rm SR}$. If $\mathcal{M}_{\rm SR}$ outputs 1, the target sample is regarded as a member; otherwise, it is classified as a non-member.

\subsection{Evaluation Results}

\mypara{Pruning Results.}
As shown in Table~\ref{pruning_result}, regardless of the pruning degree, \nameappend$_{\rm SR}$ consistently surpasses all \nameappend$_{\rm NR}$ attacks on the original model, providing compelling evidence that pruning operations lead to additional privacy leakage.
For instance, the AUC of \nameappend$_{\rm SR}$ on the 60\%-pruned VGG16 is 93.2\%, while the best \nameappend$_{\rm NR}$ achieves an AUC of 66\% on the original model.
Notably, even using the less powerful LR as the attack meta-classifier's architecture, \nameappend$_{\rm SR}$ generally outperforms \nameappend$_{\rm NR}$ on the compressed model, with the attack performance gap widening significantly when RF is employed as the attack meta-classifier.
This indicates that, irrespective of the meta-classifier's capacity, \nameappend$_{\rm SR}$ exhibits pronounced effectiveness in pruning scenarios.
For example, when targeting the 90\%-pruned VGG16 on Mini-ImageNet, \nameappend$_{\rm SR}$ improves the best TPR @0.1\% FPR from 1.3\% to 25.5\%, best AUC from 61.9\% to 88.6\%, and best balanced accuracy from 59.3\% to 79.9\%.

\mypara{Quantization Results.}
As detailed in Table~\ref{qat_result}, \nameappend$_{\rm SR}$ leverages information from both original and quantized models, surpassing all \nameappend$_{\rm NR}$'s attack performance on the original or quantized model, thereby highlighting that quantization operations indeed amplify privacy risks and demonstrating the effectiveness of our \nameappend$_{\rm SR}$.
For instance, on the Mini-ImageNet, compared to the best-performing \nameappend$_{\rm NR}$ on the original model (quantized model), \nameappend$_{\rm SR}$ achieves a significant improvement of 32.3\% (33.8\%) in AUC and 29.2\% (30.0\%) in balanced accuracy.
In addition, we also observe that the privacy leakage induced by int 8-bit QAT is substantially higher than $L_{1}$ unstructured pruning, i.e.,  at TPR @ 0.1\% FPR, the leakage is 38.9\% exacerbated than 70\%-pruned VGG16.

\mypara{Weight Clustering Results.}
Similar to pruning, although the attack performance of \nameappend$_{\rm SR}$ decreases with fewer clusters due to the proximity of member and non-member behavior as model capacity diminishes, it still outperforms all \nameappend$_{\rm NR}$ attacks on the original model (or clustered model), regardless of the number of clusters, highlighting that weight clustering imposes additional privacy leakage and the effectiveness of \nameappend$_{\rm SR}$.
Specifically, for the TPR @ 0.1\% FPR shown in Table~\ref{basic attack cluster}, \nameappend$_{\rm SR}$ presents an order of magnitude improvement compared to \nameappend$_{\rm NR}$ on the original or clustered model.

\subsection{Discussion}
In this subsection, we start by analyzing the characteristics of samples that become vulnerable to MIA after compression, then extend our evaluation to foundation models, e.g., BERT, GPT-2. In the end, we provide the results of \nameappend$_{\rm NR}$ and \nameappend$_{\rm SR}$ against MIA defenses using DP-SGD~\cite{abadi2016deep}.

\begin{table*}
\centering
\caption{The attack results of BERT-base/GPT-2 finetuned on SST-5.}
\label{bert_attack}
\scalebox{0.7}{
\begin{tabular}{c|ccccccccc}
\toprule
Attack  & \multicolumn{3}{c}{TPR @ 0.1\% FPR (\%)} & \multicolumn{3}{c}{Balanced Accuracy (\%)}  & \multicolumn{3}{c}{AUC (\%)} \\
\cmidrule(r){2-4} \cmidrule(r){5-7} \cmidrule(r){8-10}Method &original & pruned & quantized &original & pruned & quantized&original & pruned  & quantized \\ 

\toprule
    \nameappend$_{\rm NR}$~\cite{nasr2018machine} (RF)
    &\textbf{1.1}/1.1 &0.6/1.0 &0.4/0.8
    &\textbf{83.7}/77.0 &80.5/74.7 &62.4/72.6 
    &\textbf{88.7}/85.0 &87.4/82.5 &66.7/79.9\\
    \nameappend$_{\rm NR}$~\cite{yeom2018privacy}
    &0.3/1.8 &0.6/1.1 &0.2/1.0
    &69.1/67.4 &66.8/71.0 &50.3/61.4 
    &74.8/75.4 &84.8/80.1 &50.9/73.0 \\
    \nameappend$_{\rm NR}$~\cite{song2021systematic}
    &0.3/\textbf{2.9} &1.0/1.6 &0.3/0.8
    &81.8/78.6 &77.6/74.5 &63.4/71.4 
    &85.7/85.7 &91.5/87.5 &68.5/84.5 \\
    \nameappend$_{\rm NR}$~\cite{yuan2022membership}   
    &0.9/1.6 &0.5/0.6 &0.1/0.4
    &80.3/\textbf{79.6} &78.4/78.1 &63.8/73.7 
    &85.0/\textbf{85.6} &81.0/81.7 &68.0/81.7 \\
    \midrule
    \nameappend$_{\rm SR}$ 2 (LR)
    &- &0.6/\textbf{2.4} &0.9/\textbf{1.9}
    &- &\textbf{85.6}/\textbf{80.7} &75.6/ \textbf{76.7}
    &- &\textbf{91.8}/\textbf{87.7} &81.9/\textbf{84.5}\\
    \nameappend$_{\rm SR}$ 2 (RF)
    &- &\textbf{1.6}/0.9 &\textbf{1.1}/1.5
    &- &84.7/76.6 &\textbf{78.8}/71.7 
    &- &91.0/84.4 &\textbf{85.2}/82.9 \\
\bottomrule
\end{tabular}
}
\end{table*}

\begin{figure}[t]
    \centering
    \begin{minipage}{0.48\linewidth}  
        \centering
        \includegraphics[width=\textwidth]{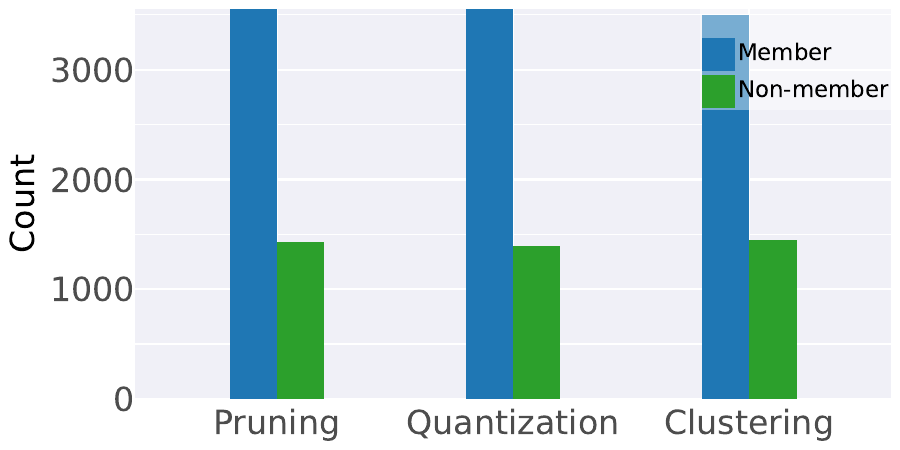}
        \subcaption{Mini-ImageNet}\label{fig:number_mini}
    \end{minipage}\hspace{1pt}  
    \begin{minipage}{0.48\linewidth}  
        \centering
        \includegraphics[width=\textwidth]{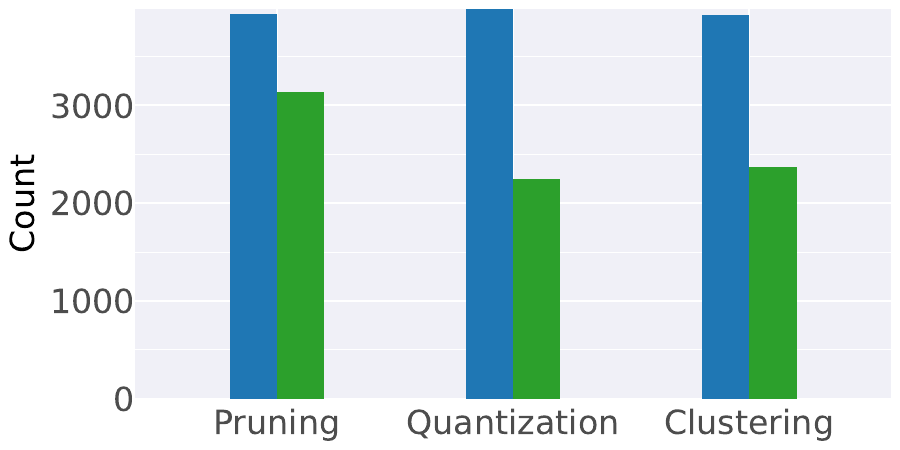}
        \subcaption{CIFAR-10}\label{fig:number_CIFAR10}
    \end{minipage}
    \caption{Number of members and non-members that transitioned from attack failure to success after compression. For Mini-ImageNet, the number of members and non-members is 8000, and for CIFAR-10, it is 13500. The pruning level is 0.9, the number of clusters is 8.}
    \label{fig:difference}
\end{figure}

\begin{figure}[t]
    \centering
    \includegraphics[trim=0 0 0 0,clip,width=0.3\textwidth]{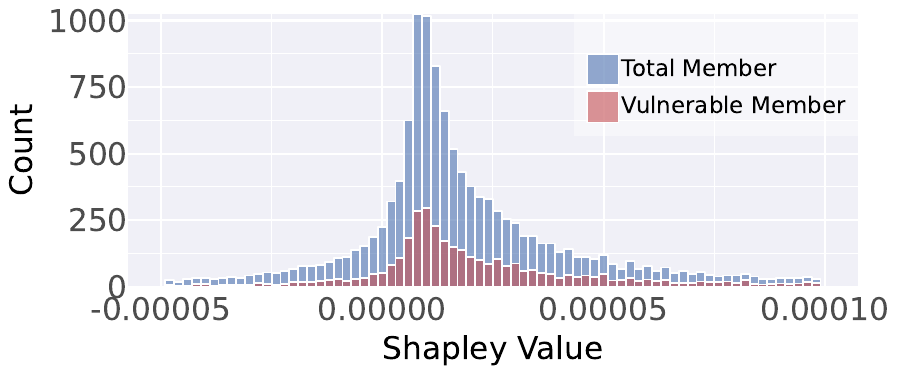}
    \caption{KNN-Shapley values of total member and vulnerable member in pruning scenario of CIFAR-10.}
    \label{fig:shapley}
\end{figure}

\begin{table*}[]
\centering
\caption{Attack performance against 70\%-pruned FCN trained on Location with DP-SGD ( $\sigma$ = 0.2 and $\sigma$ = 0.5). The values in parentheses represent the attack performance on the original FCN.}
\label{dp}
\scalebox{0.7}{
\begin{threeparttable}
\begin{tabular}{c|ccccccccc}
\toprule
Attack  & \multicolumn{3}{c}{TPR @ 0.1\% FPR (\%)} & \multicolumn{3}{c}{Balanced Accuracy (\%)}  & \multicolumn{3}{c}{AUC (\%)} \\
\cmidrule(r){2-4} \cmidrule(r){5-7} \cmidrule(r){8-10}Method &no defense & $\sigma$ = 0.2 & $\sigma$ = 0.5 &no defense &$\sigma$ = 0.2 & $\sigma$= 0.5 &no defense &$\sigma$ = 0.2 & $\sigma$ = 0.5 \\ 
    
\toprule

    \nameappend$_{\rm NR}$~\cite{shokri2017membership} 
    &2.8 (6.0)&0.5 (1.5) &0.2 (1.4)
    &81.2 (80.1)&64.9 (67.6) &59.7 (63.4)
    &87.9 (91.7)&69.4 (73.5) &63.5  (67.0)\\
    \nameappend$_{\rm NR}$~\cite{nasr2018machine}
    &3.2 (3.6)&0.4 (1.6) &0.0 (1.6)
    &80.7 (79.3)&64.1 (67.3) &60.0 (62.5)
    &87.2 (91.7)&69.0 (73.3) &63.3 (66.9)\\
    \nameappend$_{\rm NR}$~\cite{yeom2018privacy}
    &0.2 (0.4)&0.0 (0.1)&0.0 (0.1)
    &84.7 (81.7)&64.3 (69.4)&59.3 (63.4)
    &88.1 (89.5)&68.5 (74.1)&62.0 (66.9)\\
    \nameappend$_{\rm NR}$~\cite{song2021systematic}
    &0.2 (0.3)&0.0 (0.2)&0.0 (0.2)
    &87.6 (84.6)&69.6 (70.7)&64.6 (66.1)
    &89.8 (90.6)&74.0 (77.5)&68.2 (71.2)\\
    \midrule
    \nameappend$_{\rm SR}$ 1 
    &\textbf{9.3} & 2.8 &\textbf{0.3}
    &88.2 &69.8 &63.8 
    &93.2 &76.2 &68.6 \\ 
    \nameappend$_{\rm SR}$ 2 
    &9.0 &\textbf{4.8} &0.2
    &\textbf{88.9} &\textbf{72.0}  &\textbf{66.1}
    &\textbf{93.7} &\textbf{78.8} &\textbf{72.0} \\ 
\bottomrule
\end{tabular}
\begin{tablenotes} 
    \item The structure of the attack meta-classifier is all based on the RF.
\end{tablenotes} 
\end{threeparttable}}
\end{table*}

\mypara{Which Data Samples are Vulnerable?}
In the above experiments, we have confirmed that pruning, quantization, and weight clustering all lead to an increased risk of privacy leakage. 
Here, we further identify which samples amplify this leakage post-compression and analyze their characteristics. 

To begin with, we focus on samples that transitioned from attack inference failure using \nameappend$_{\rm NR}$ on the original model (measured by ~\cite{nasr2018machine}), to success after compression when employing \nameappend$_{\rm SR}$. 
Firstly, as revealed in Figure \ref{fig:difference}, most of these samples are members. Since members are more private than non-members, this exposes a critical vulnerability in model compression that enables adversaries to extract more sensitive information. 
Secondly, to assess the importance of these vulnerable member samples within the entire member set, we followed the approach outlined in \cite{wen2024understanding}, utilizing KNN-Shapley \cite{jia2019efficient} to compute the importance of both all members and vulnerable members. 
As larger Shapley values represent higher importance, Figure \ref{fig:shapley} illustrates these vulnerable members are relatively more important within the overall member set. These critical findings highlight that model compression not only increases the number of member samples effectively inferred, but also amplifies the exposure of high-value members.

\begin{figure}[t]
\centering
    \includegraphics[scale=0.22]{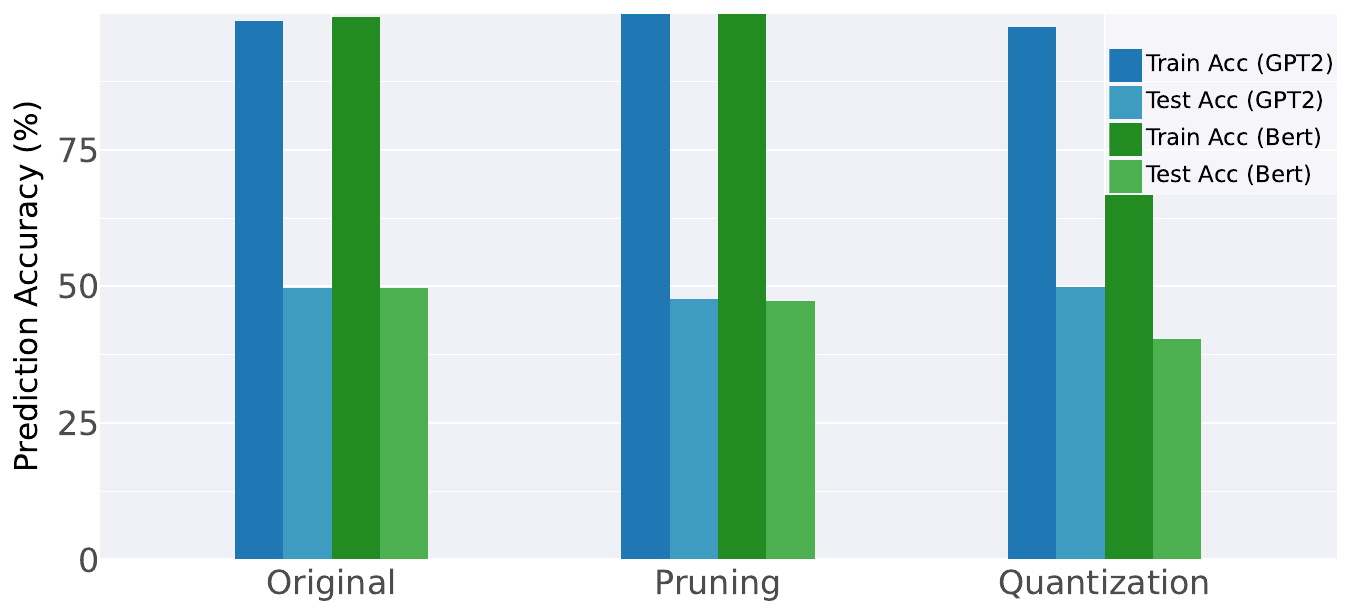} 
    \caption{The prediction accuracy of the original and compressed BERT/GPT-2 fine-tuned on SST-5.}
    \label{fig:bert_pre}
\end{figure}
\mypara{Attack Against Foundation Models.}
We quantify the membership leakage during the high-risk fine-tuning phase of foundation models~\cite{zhou2024comprehensive}, which raises unique concerns due to the processing of sensitive proprietary data, in contrast to the pre-training phase operates on public corpora.
Furthermore, we focus on the two most commonly studied compression operations for foundation models: pruning and quantization. The detailed settings are as follows:

\noindent \textit{Pruning.}
We utilize a simple and effective \textit{layer pruning}~\cite{men2024shortgpt} tailored for foundation models. As discussed in Section~\ref{sec:related}, we assess the importance of each layer using block influence and discard the less important ones \cite{men2024shortgpt}.
    
\noindent\textit{Quantization.} 
High-performing QAT typically requires additional fine-tuning, which can be computationally expensive for foundation models because of their massive parameters. Consequently, in this work, we evaluate based on PyTorch's \textit{dynamic quantization}—a form of post-training quantization—eliminating fine-tuning and directly quantizing the foundation models into int 8-bit format.

In our experiments, we fine-tune BERT-base/GPT-2 for 30 epochs on downstream tasks: SST-5~\cite{socher2013recursive}. To consider a balance between model accuracy and compression, we remove six unimportant layers from BERT-base and one from GPT-2, and apply dynamic quantization to convert the fine-tuned model to 8-bit integer precision.
Figure~\ref{fig:bert_pre} presents the prediction accuracy of pruned and quantized BERT/GPT-2. While BERT exhibits notable accuracy drop after quantization, other configurations maintain comparable performance to the original model, with only marginal declines.
Importantly, as shown in Table~\ref{bert_attack}, pruning sometimes weakens \nameappend${\rm NR}$'s effectiveness, whereas our \nameappend${\rm SR}$ still achieves the highest performance. This suggests that removing redundant layers in the foundation model inadvertently leaks privacy. In addition, quantization degrades all MIA capabilities, particularly for \nameappend$_{\rm NR}$.
We attribute this to the absence of retraining after dynamic quantization, which prevents the model from recovering the memory of member-specific features, especially in BERT, where we observe a notable 33\% drop in prediction accuracy on members.

\mypara{Attack Against DP-SGD.}
Differential privacy~\cite{dwork2006calibrating} is a widely adopted defense mechanism for mitigating privacy leakage risks. Following~\cite{li2024seqmia}, we implement DP-SGD via the Opacus toolkit with privacy parameters ($\delta$ = 1e-5, $C$ = 1). 
However, stronger defenses typically lead to a significant drop in model utility, which could be unacceptable in practice.
To carefully consider the trade-off between the defense level and model accuracy, we choose two noise multipliers $\sigma \in \{0.2, 0.5\}$, where a larger $\sigma$ provides stronger protection.
Table~\ref{dp} presents the attack performance under DP-SGD, evaluated on 70\%-pruned FCN on Location. 

We observe that DP-SGD offers an effective defense against all MIAs, with attack performance gradually decreasing as defense strength increases. 
Notably, \nameappend$_{\rm SR}$ consistently achieves the best attack performance, compared to \nameappend$_{\rm NR}$ target either the original or compressed model. This highlights that the compression operation still leads to additional leakage even after deploying the defense strategy. 
However, as defensive capability strengthens, this additional leakage decreases because the gap in attack effectiveness between \nameappend$_{\rm SR}$ and \nameappend$_{\rm NR}$ (target original model) narrows.
For instance, when \nameappend$_{\rm NR}$ selects ~\cite{nasr2018machine}, the balanced accuracy gap decreases from 9.6\% (no defense) to 4.7\% ($\sigma$=0.2) and 3.6\% ($\sigma$=0.5).

\begin{table}[t]
\centering
\caption{Attack performance against  80\%-pruned ResNet-18 trained on CIFAR-10 with different overfitting levels (values in parentheses represent the performance on the original ResNet18).}
\label{overfitting}
\scalebox{0.80}{
    \begin{threeparttable}
    \begin{tabular}{c|cccc}
    \toprule
    Attack  & \multicolumn{2}{c}{Balanced Accuracy (\%)} & \multicolumn{2}{c}{AUC (\%)}\\
    \cmidrule(r){2-3} \cmidrule(r){4-5} Method & $L_o$ = 29\% & $L_o$ = 10\% & $L_o$ = 29\% & $L_o$ = 10\% \\ 
    
    \midrule
    \nameappend$_{\rm NR}$~\cite{shokri2017membership}   
    &68.5 (65.6) &57.8 (60.3) &74.5 (71.4) & 60.2 (64.2) \\
    \nameappend$_{\rm NR}$~\cite{yeom2018privacy}
    &68.8 (65.5) &60.5 (61.0) &71.9 (69.1) & 61.6 (62.2) \\
    \midrule
    \nameappend$_{\rm SR} 2$ 
    &\textbf{72.2} &\textbf{61.7} &\textbf{77.2} &\textbf{65.4}\\               
    \bottomrule
\end{tabular}
\begin{tablenotes} 
\footnotesize 
    \item When $L_o$ = 29\%, the training accuracy is 99\%, the test accuracy is 70\%, and the training dataset contains 135000 samples, following~\cite{yuan2022membership}.  When $L_o$ = 10\%, the training accuracy is 99\%, the test accuracy is 89\%, and the training dataset contains 20000 samples.
    \item Attack meta-classifier adopts the RF architecture.
\end{tablenotes} 
\end{threeparttable}}
\end{table}

\subsection{Ablation Study}
We conduct experiments to examine the influence of several key factors (i.e., overfitting level of the victim model, amount of data used for fine-tuning) on \nameappend$_{\rm SR}$ performance.
Because of limited space, an analysis of dataset effects is provided in Appendix~\ref{appendix:victim_dataset}.
All subsequent experiments employ pruning as the representative compression operation.

\mypara{Overfitting Level of the Victim Model}.
It is widely recognized that membership inference attacks are closely related to the overfitting level of the victim model.
Following~\cite{li2024seqmia}, we quantify the overfitting level by measuring the gap between training and testing accuracy and regulate it by varying the size of the training dataset.
Specifically, we consider two distinct levels $L_o \in \{29\%, 10\%\}$ to explore their influence on MIA performance.

As described in Table~\ref{overfitting}, all MIAs demonstrate weaker performance against models with low overfitting, which conclusion aligns with~\cite{li2024seqmia}.
In addition, we observe that, under low-overfitting conditions, the 80\%-pruned model is less susceptible to \nameappend$_{\rm NR}$ than the original model, which aligns with our earlier findings in Section~\ref{NR_E}.
However, as overfitting increases, this trend reverses. Notably, \nameappend$_{\rm SR}$ is always superior over \nameappend$_{\rm NR}$ on the original/compressed model, with the performance gap becoming particularly pronounced in high-overfitting scenarios.

\mypara{Amount of Data Used for Fine-tuning. }
Fine-tuning the compressed model using the original training dataset is a crucial step in many compression operations.
In the experiments above, we utilized the entire training dataset for fine-tuning. However, it is more practical to fine-tune the model using a selected subset of the original training dataset, as this can considerably reduce fine-tuning time. This subset, denoted as $\mathcal{D}_{f}$ ( $\lvert \mathcal{D}_{f} \rvert= {N}_{f}$), typically contains more valuable and sensitive data, while the remainder of the original training dataset, not used for fine-tuning, is denoted as $\mathcal{D}_{nf}$. 
Then, we explore the impact of ${N}_{f}$ by fine-tuning the compressed model using three portions (90\%, 50\%, and 10\%) of the original training dataset and measuring the attack performance on both $\mathcal{D}_{f}$ and $\mathcal{D}_{nf}$. 

Table~\ref{fine-tuning} reports that the privacy leakage for $\mathcal{D}_{f}$ is higher than for $\mathcal{D}_{nf}$, and this gap becomes more pronounced as ${N}_{f}$ decreases. 
This stems from the exclusion of $\mathcal{D}_{nf}$ during fine-tuning, which reduces model accuracy on this dataset and causes its behavior to resemble those for non-members.
Moreover, the MIA performance on $\mathcal{D}_{f}$ gradually enhances as ${N}_{f}$ decrease. 
We hypothesize that this occurs because, as ${N}_{f}$ decreases, the model learns more refined features from $\mathcal{D}_{f}$, while its generalization capability to non-members deteriorates, as depicted in Table~\ref{acc_f}, this leads to an increasing disparity between $\mathcal{D}_{f}$ and non-members. 
Thus, fine-tuning using a subset of the training dataset exacerbates privacy leakage risks for the more valuable data used during fine-tuning. 

\begin{table}[t]
\centering
\caption{Attack performance on $\mathcal{D}_{f}$/$\mathcal{D}_{nf}$ with different ${N}_{f}$ (80\%-pruned ResNet18+CIFAR-10).}
\label{fine-tuning}
\scalebox{0.7}{
    \begin{tabular}{lcccccc}
    \toprule
    Attack& \multicolumn{3}{c}{Balanced Accuracy} & \multicolumn{3}{c}{AUC} \\
    \cmidrule(r){2-4} \cmidrule(r){5-7} 
    Method& 90\% & 50\% & 10\% & 90\% & 50\% & 10\% \\ 
    \midrule
    \nameappend$_{\rm NR}$ ~\cite{shokri2017membership} 
    & 59.2/57.2 & 59.6/55.6 & 61.8/56.6 
    & 62.3/60.0 & 62.9/58.2 & 65.8/58.9 \\
    \nameappend$_{\rm NR}$ ~\cite{nasr2018machine}   
    & 59.6/57.4 & 60.2/56.0 & 62.1/56.8 
    & 62.9/60.5 & 63.3/59.0 & 66.2/59.6 \\
    \nameappend$_{\rm SR} 2$        
    & 61.8/61.4 & 62.2/60.9 & 64.2/61.0 
    & 65.4/65.0 & 65.7/64.6 & 67.9/64.8 \\
    \bottomrule
\end{tabular}}
\end{table}

\section{\nameappend$_{\rm MR}$}
\label{sec:MR}
Through \nameappend$_{\rm SR}$, which leverages a single compressed model as a reference, we conclude that compression operations indeed lead to additional privacy leakage. 
More interestingly, in this section, based on the intuition that different compressed models leak privacy in slightly different ways, we demonstrate that when multiple compressed models are utilized as references---referred to as \nameappend$_{\rm MR}$---the privacy leakage induced by compression is further exacerbated. 
We begin by considering two distinct adversarial settings---whether the original model is accessible or not. 
Then, we present the attack methodology, and conclude with the evaluation results and ablation studies.

\subsection{Adversarial Knowledge}
\label{threat_MR}

Here, we assume two different threat models, gradually limiting the adversary's knowledge to show the broader attack scenarios.

\mypara{Adversary 1:} 
Consistent with the threat model described in Section~\ref{threat}, the adversary is assumed to have black-box access to the original model as well as its multiple compressed versions. Moreover, they also know the sparsity level or model size associated with each compressed model.

\mypara{Adversary 2:}
Compared to Adversary 1, Adversary 2 adopts a stricter threat model.
Following the setup in~\cite{yuan2022membership}, we assume that the adversary has black-box access only to multiple compressed models, with {\it no access} to the original model.
All other conditions are the same as Adversary 1.

\begin{figure}[t]
    \centering
    \includegraphics[trim=0 0 0 0,clip,width=0.4\textwidth]{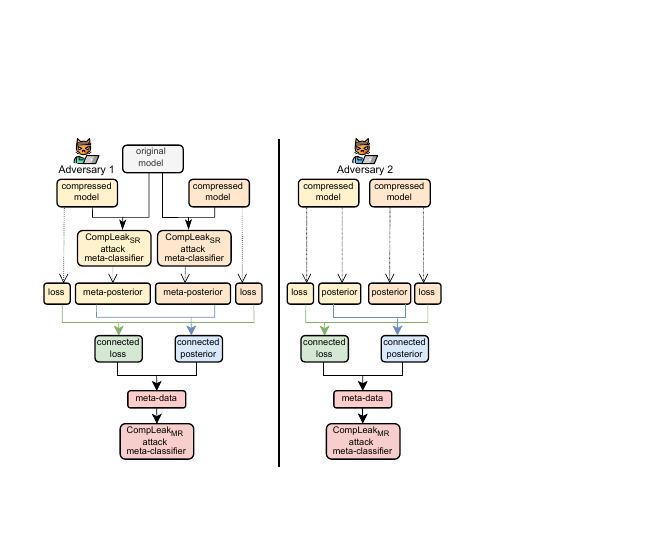} 
    \caption{General attack pipeline of \nameappend$_{\rm MR}$, using two compressed models for example. }
    \label{fig:MR}
\end{figure}
\subsection{Attack Methodology}

When the reference model is expanded from a single compressed model to multiple compressed models, we observe that different compressed versions leak privacy differently, revealed by both the loss and the membership inference results from the \nameappend$_{\rm SR}$ attack meta-classifier.
Therefore, to implement \nameappend$_{\rm MR}$, the adversary aggregates the leaked information by extracting and concatenating the losses and \nameappend$_{\rm SR}$ meta-posteriors through each compressed model and \nameappend$_{\rm SR}$ attack meta-classifiers separately, then stacks them to train an MLP-based \nameappend$_{\rm MR}$ attack meta-classifier for membership inference.

Similar to \nameappend$_{\rm SR}$, performing \nameappend$_{\rm MR}$ involves five key stages: shadow model training, loss concatenation, posterior concatenation, attack meta-classifier training, and attack meta-classifier membership inference. We present the detailed attack pipeline of \nameappend$_{\rm MR}$ in Figure~\ref{fig:MR}. The following description is based on the scenario of Adversary 1, any differences for Adversary 2 will be specified separately.

\mypara{Shadow Model Training.}
In the context of a multi-compressed model scenario, there are three types of shadow models. More concretely, the adversary first trains a shadow original model $\mathcal{M}_{o}^{s}$ on the shadow training set $\mathcal{D}_{\rm train}^{s}$, then applies the compression algorithm to generate the corresponding shadow compressed models $\mathcal{M}_{c_{1}}^{s},\mathcal{M}_{c_{2}}^{s},\dots,\mathcal{M}_{c_{n}}^{s}$. Finally, the adversary conducts \nameappend$_{\rm SR}$ on each shadow compressed model, training a set of shadow \nameappend$_{\rm SR}$ attack meta-classifiers $\mathcal{M}_{\rm SR_{1}}^{s},\mathcal{M}_{\rm SR_{2}}^{s},\dots,\mathcal{M}_{\rm SR_{n}}^{s}$.
Notably, Adversary 2 is unable to get $\mathcal{M}_{\rm SR_{i}}^{s}$.

\mypara{Posterior Concatenation.}
We first present the design rationale, followed by the detailed implementation.

\textit{Design Rationale.}
Building on \nameappend$_{\rm SR}$, we observe that the membership inference results from $\mathcal{M}_{\rm SR_{1}}^{s},\mathcal{M}_{\rm SR_{2}}^{s},\dots,\mathcal{M}_{\rm SR_{n}}^{s}$ on the same target sample exhibit striking differences. For example, when the $\mathcal{M}_{\rm SR_{i}}^{s}$ are all based on a random forest structure, the inference results targeting an 80\%-pruned VGG16 on Mini-ImageNet differ by approximately 22\% from targeting a 90\%-pruned VGG16.
This can be explained by different compression degrees exerting varying influences on the same sample, and through $\mathcal{M}_{\rm SR_{1}}^{s},\mathcal{M}_{\rm SR_{2}}^{s},\dots,\mathcal{M}_{\rm SR_{n}}^{s}$, we can effectively capture these subtle differences.

\begin{figure}[t]
    \begin{minipage}{0.24\linewidth}  
        \centering
        \includegraphics[width=\textwidth]{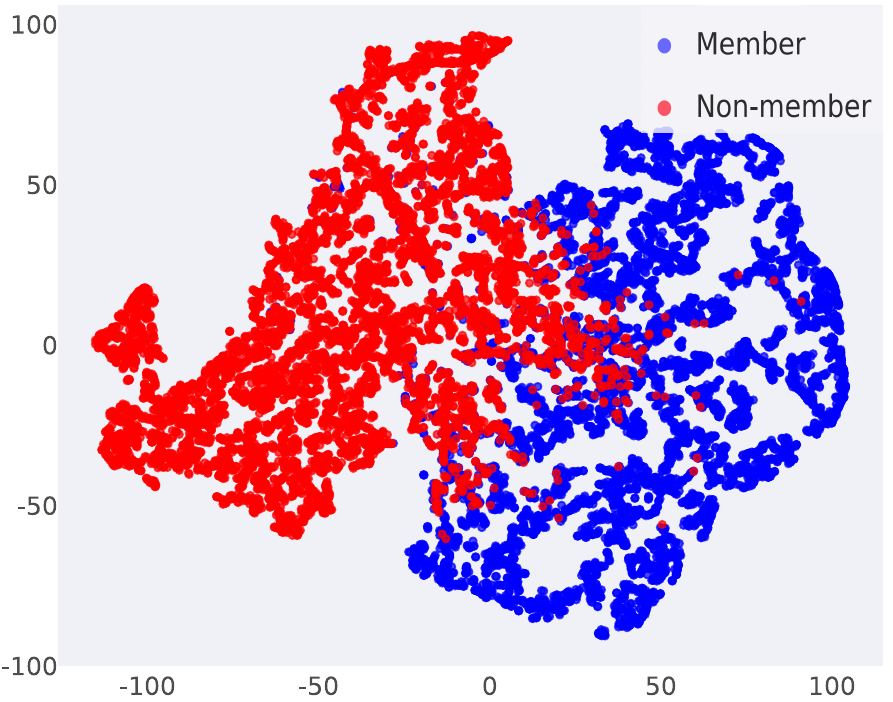}
        \subcaption{\small $\mathcal{P}_{SR}$ }\label{fig:attack_score}
    \end{minipage} 
    \begin{minipage}{0.24\linewidth}  
        \centering
        \includegraphics[width=\textwidth]{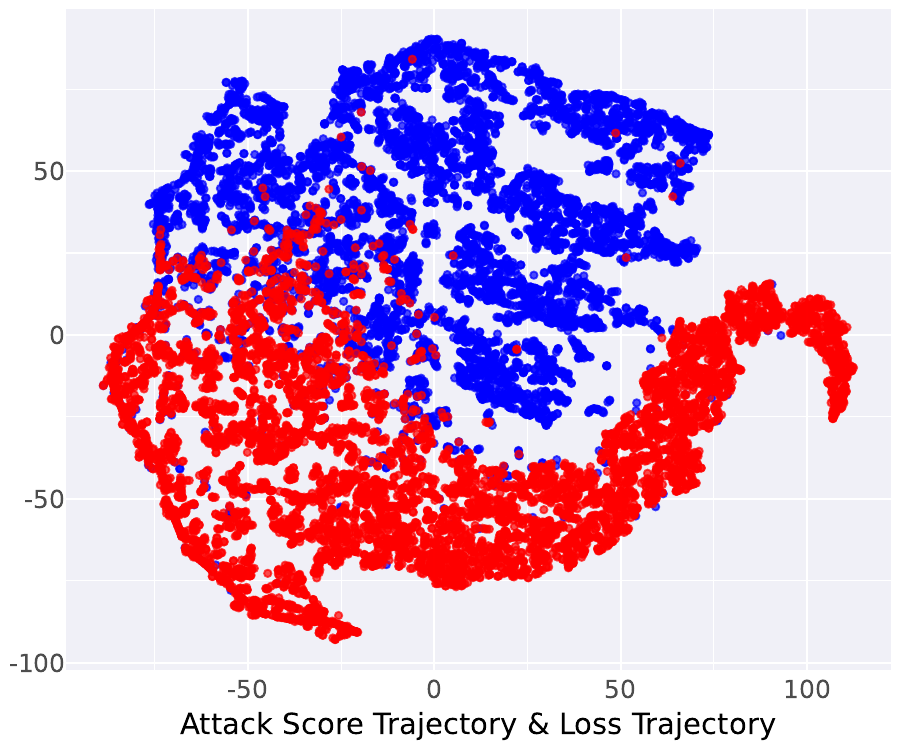}
        \subcaption{\small $\mathcal{P}_{SR}$ \& $\mathcal{L}$}\label{fig:loss_attack_score}
    \end{minipage}  
    \begin{minipage}{0.24\linewidth}  
        \centering
        \includegraphics[width=\textwidth]{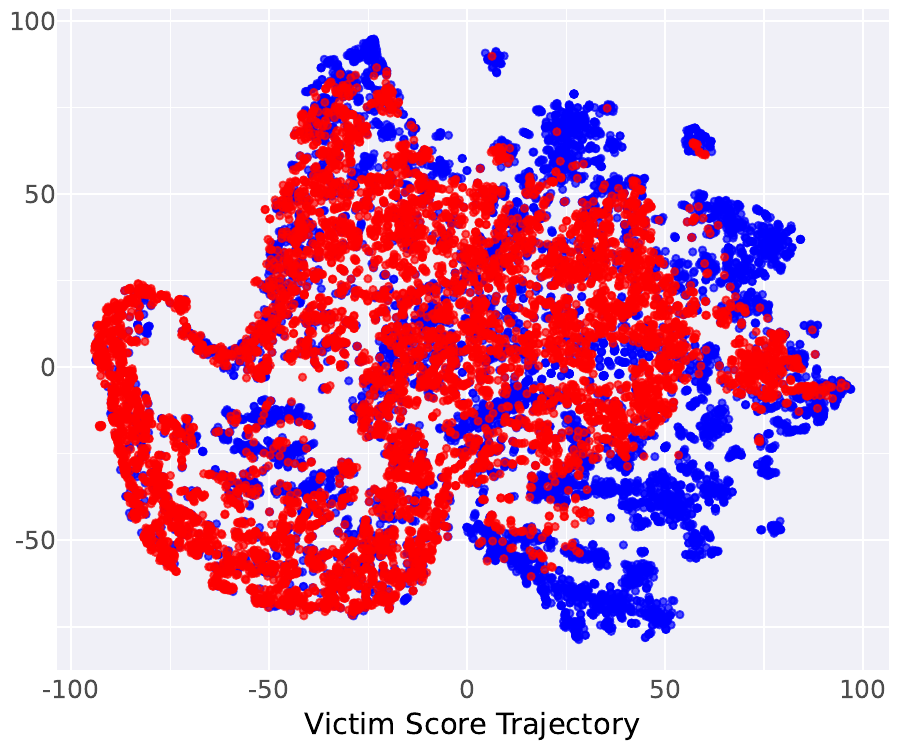}
        \subcaption{\small $\mathcal{P}_{c}$ }\label{fig:victim_score}
    \end{minipage}
    \begin{minipage}{0.24\linewidth}  
        \centering
        \includegraphics[width=\textwidth]{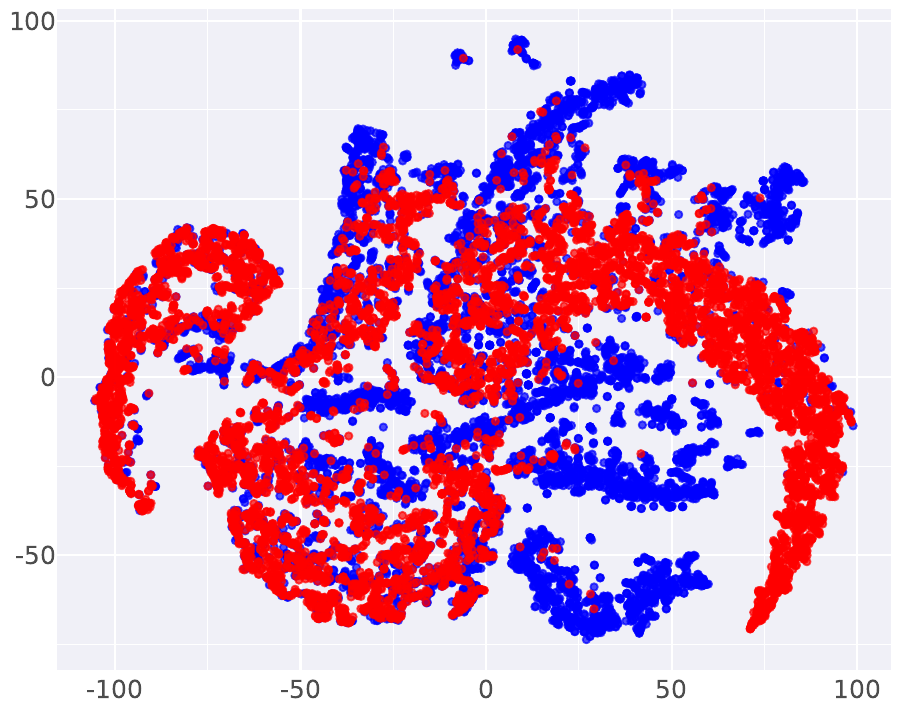}
        \subcaption{\small $\mathcal{P}_{c}$ \& $\mathcal{L}$}\label{fig:loss_victim_score}
    \end{minipage} 
    \caption{Using t-SNE to visualize the four distributions. $\mathcal{P}_{SR}$, $\mathcal{P}_{c}$ refers to posterior concatenation for Adversary 1 and Adversary 2, and $\mathcal{L}$ refers to loss concatenation.}
    \label{fig:t-sne}
\end{figure}

\textit{Implementation.}
Due to the differing knowledge of the two adversaries, the methods for obtaining the posteriors also differ.
\begin{itemize}[leftmargin=*]
    \item Adversary 1. Because the original model is accessible under this assumption, the adversary can train $\mathcal{M}_{\rm SR_{1}}^{s},\mathcal{M}_{\rm SR_{2}}^{s},\dots,\mathcal{M}_{\rm SR_{n}}^{s}$ through \nameappend$_{\rm SR}$ during the Shadow Model Training step.
    Based on design rationale, the adversary queries \nameappend$_{\rm SR}$ meta-classifiers $\mathcal{M}_{\rm SR_{1}}^{s},\mathcal{M}_{\rm SR_{2}}^{s},\dots,\mathcal{M}_{\rm SR_{n}}^{s}$ for each sample from $\mathcal{D}_{s}$ to get \nameappend$_{\rm SR}$ meta-posteriors $\mathcal{P}_{\rm SR_{1}}^{s},\mathcal{P}_{\rm SR_{2}}^{s},\dots,\mathcal{P}_{\rm SR_{n}}^{s}$.
    \item Adversary 2. Due to the lack of access to the original model, the adversary is unable to acquire $\mathcal{M}_{\rm SR_{1}}^{s},\mathcal{M}_{\rm SR_{2}}^{s},\dots,\mathcal{M}_{\rm SR_{n}}^{s}$. Instead, the posteriors here are derived by directly querying $\mathcal{M}_{c_{1}}^{s}, \mathcal{M}_{c_{2}}^{s}, \dots, \mathcal{M}_{c_{n}}^{s}$ for each sample from $\mathcal{D}_{s}$, denoted as $\mathcal{P}_{c_{1}}^{s},\mathcal{P}_{c_{2}}^{s},\dots,\mathcal{P}_{c_{n}}^{s}$.
\end{itemize}
Finally, the adversary performs the concatenation operation on $\mathcal{P}_{\rm SR_{1}}^{s},\mathcal{P}_{\rm SR_{2}}^{s},\dots,\mathcal{P}_{\rm SR_{n}}^{s}$ ($\mathcal{P}_{c_{1}}^{s},\mathcal{P}_{c_{2}}^{s},\dots,\mathcal{P}_{c_{n}}^{s}$) to generate $\mathcal{P}_{\rm SR}^{s}$ ($\mathcal{P}_{c}^{s}$), i.e., $\mathcal{P}_{\rm SR}^{s} = \mathcal{P}_{\rm SR_{1}}^{s} \parallel \mathcal{P}_{\rm SR_{2}}^{s} \parallel \dots \parallel \mathcal{P}_{\rm SR_{n}}^{s}$.
We visualize $\mathcal{P}_{\rm SR}^{s}$ using t-SNE~\cite{van2008visualizing} in Figure~\ref{fig:attack_score}, a clear boundary is observed between members and non-members compared to $\mathcal{P}_{c}^{s}$ in Figure~\ref{fig:victim_score}. 

\begin{figure}[t]
    \centering
    \includegraphics[trim=0 0 0 0,clip,width=0.25\textwidth]{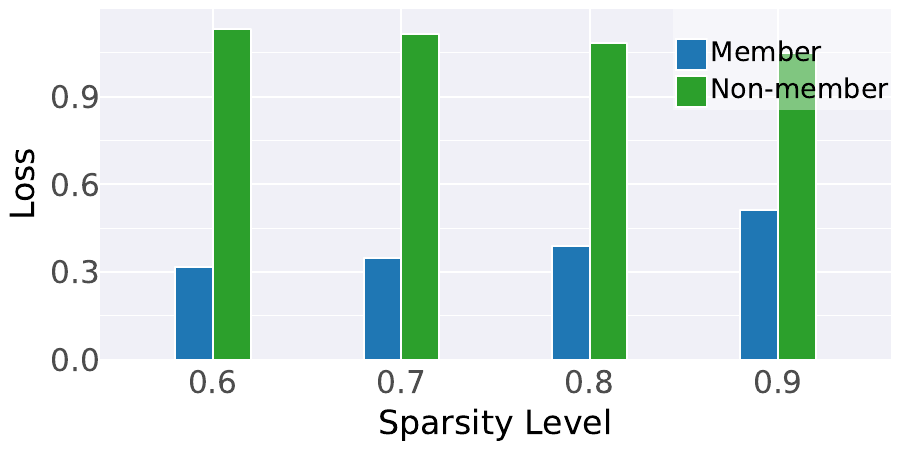}
    \caption{Average loss of members and non-members under varying sparsity levels for the VGG16 pruned on the Mini-ImageNet. }
    \label{fig:loss}
\end{figure}

\mypara{Loss Concatenation.}
We begin with the design rationale, and then describe the implementation.

\textit{Design Rationale.}
We find that as the compression degree increases, the evolution of loss calculated from compressed models on the same target sample reveals disparities between members and non-members, both in direction and magnitude. 
For clarity, we use pruning as an example to explain.
Specifically, as illustrated in Figure~\ref{fig:loss}, the loss for members rises with higher sparsity levels, while the loss for non-members fluctuates. 
This can be attributed to the sparse double descent found in~\cite{hoefler2021sparsity}, as model capacity decreases, non-member accuracy declines while member accuracy remains stable, however, after a critical sparsity point, member accuracy drops and non-member accuracy rises before declining again.

\textit{Implementation.}
The adversary feeds each data sample from $\mathcal{D}_{s}$ to $\mathcal{M}_{c_{1}}^{s}, \mathcal{M}_{c_{2}}^{s}, \dots, \mathcal{M}_{c_{n}}^{s}$, calculating a series of loss values $\mathcal{L}_{1}^{s},\mathcal{L}_{2}^{s},\dots,\mathcal{L}_{n}^{s}$.
The loss \( \mathcal{L}_i^{s} \) for each shadow compressed model \( \mathcal{M}_{c_{i}}^{s} \) is calculated through the cross-entropy:
$\mathcal{L}_i^{s} = - \sum_{k=1}^{C} y_k \log(p_k^{(i)})$.
Where \( C \) is the number of classes, \( y_k \) is the true label for class \( k \) (one-hot encoded), and \( p_k^{(i)} \) is the predicted posterior probability for class \( k \) generated from the model \( \mathcal{M}_{c_{i}}^{s} \).
Finally, the adversary concatenates each loss to obtain $\mathcal{L}_{s}$, i.e., $\mathcal{L}_{s} = \mathcal{L}_{1}^{s} \parallel \mathcal{L}_{2}^{s} \parallel \dots \parallel \mathcal{L}_{n}^{s}$.
As illustrated in Figures~\ref{fig:loss_attack_score} and ~\ref{fig:loss_victim_score}, incorporating the $\mathcal{L}_{s}$---whether for $\mathcal{P}_{\rm SR}^{s}$ or $\mathcal{P}_{c}^{s}$---further enhances the ability to distinguish between members and non-members.


\mypara{Attack Meta-classifier Training.}
The adversary constructs the attack binary training dataset $\mathcal{D}_{\rm train}^{attack}$ by stacking $\mathcal{L}_{s}$ and $\mathcal{P}_{\rm SR}^{s}$ (or $\mathcal{P}_{c}^{s}$ for Adversary 2), labeling the stacked data as 1 (member) if it comes from $\mathcal{D}_{\rm train}^{s}$, and 0 (non-member) otherwise.
Subsequently, the adversary trains an MLP-based \nameappend$_{\rm MR}$ meta-classifier $\mathcal{M}_{\rm MR}$ on $\mathcal{D}_{\rm train}^{attack}$ to perform membership inference.

\mypara{Attack Meta-classifier Membership Inference.}
The attacker can ultimately conduct MIA on each given target sample following these steps: First, the adversary performs \nameappend$_{\rm SR}$ on each victim's compressed models to train the victim  \nameappend$_{\rm SR}$ attack meta-classifiers $\mathcal{M}_{\rm SR_{1}}^{v},\mathcal{M}_{\rm SR_{2}}^{v},\dots,\mathcal{M}_{\rm SR_{n}}^{v}$ (for Adversary 2, this step is not required). 
Then, the target sample is subjected to loss concatenation and posterior concatenation to obtain $\mathcal{L}_{v}$ and $\mathcal{P}_{\rm SR}^{v}$ (or $\mathcal{P}_{c}^{v}$ for Adversary 2), respectively. Finally, the adversary stacks $\mathcal{L}_{v}$ and $\mathcal{P}_{\rm SR}^{v}$ (or $\mathcal{P}_{c}^{v}$), feeding them into $\mathcal{M}_{\rm MR}$ to predict the sample's membership status, i.e., 1 (member) or 0 (non-member).

\subsection{Evaluation}

\mypara{Experimental Setup.}
For the selection of compression degrees in multiple compressed models, we choose pruned models with sparsity levels $L=\{0.6,0.7,0.8,0.9\}$, clustered models with cluster centers $N=\{4,8,16\}$.

\mypara{Results for Adversary 1.}
Table~\ref{composite attack (Adversary 1)} depicts the performance of \nameappend$_{\rm MR}$ (Adversary 1) and highlights the improvements achieved compared to the best performance of \nameappend$_{\rm SR}$ on a single compressed model. 
Encouragingly, we observe that the \nameappend$_{\rm MR}$ clearly demonstrates superior performance compared to \nameappend$_{\rm SR}$ in TPR @ 0.1\% FPR.
Notably, when multiple compressed models are derived from three compression operations (eight models in total), all evaluation metrics show exceptionally high values, with the AUC approaching 100\%.
These results provide strong evidence that multiple compressed models leak significantly more information compared to a single compressed model.

\begin{table}[t] 
\caption{Average performance of \nameappend$_{\rm MR}$ on Mini-ImageNet under Adversary 1 (five repetitions)}.
\label{composite attack (Adversary 1)}
\scalebox{0.7}{
\begin{threeparttable}
\renewcommand{\arraystretch}{1.2}
    \begin{tabular}{c|ccc}
    \hline
    Operation & TPR @ 0.1\% FPR (\%) & Balanced Accuracy (\%)& AUC (\%)\\ 
    \hline
    Pruning     
    & 61.3 (19.2\% $\uparrow$) 
    & 94.8 (10.5\% $\uparrow$) 
    & 98.9 (5.7\% $\uparrow$) \\
    Clustering 
    & 72.7 (5.5\% $\uparrow$)
    & 95.1 (1.9\% $\uparrow$)
    & 99.0 (0.3\% $\uparrow$)\\
    Three Operations 
    & 95.7 (14.7\% $\uparrow$)
    & 98.8 (5.6\% $\uparrow$)
    & 99.9 (1.2\% $\uparrow$)\\ 
    \bottomrule
    \end{tabular}
\begin{tablenotes} 
    \item In parentheses is the improvement over the optimal result when \nameappend$_{\rm SR}$ attacking a single compressed model in the respective compression scenario.
\end{tablenotes} 
\end{threeparttable}
}
\end{table}

\begin{table}[t]
\caption{Average performance of \nameappend$_{\rm MR}$ on Mini-ImageNet under Adversary 2 (five repetitions).}
\label{composite attack (Adversary 2)}
\scalebox{0.7}{
\begin{threeparttable}
\renewcommand{\arraystretch}{1.2}
    \begin{tabular}{c|ccc}
    \hline
    Operation & TPR @ 0.1\% FPR (\%) & Balanced Accuracy (\%)& AUC (\%)\\ 
    \hline
    Pruning     
    & 2.6 (1.8\% $\uparrow$) 
    & 71.3 (9.6\% $\uparrow$) 
    & 77.6 (11.2\% $\uparrow$) \\
    Clustering 
    & 1.1 (0.2\% $\downarrow$)
    & 65.6 (3.9\% $\uparrow$)
    & 70.6 (6.3\% $\uparrow$)\\
    Three Operations 
    & 8.8 (7.5\% $\uparrow$)
    & 71.4 (9.7\% $\uparrow$)
    & 80.6 (14.2\% $\uparrow$)\\ 
    \hline
    \end{tabular}
    \begin{tablenotes} 
        \item In parentheses is the improvement over the optimal result when SAMIA~\cite{yuan2022membership} attacking a single compressed model in the respective compression scenario.
    \end{tablenotes} 
\end{threeparttable}
}
\end{table}

\mypara{Results for Adversary 2.}
As previously mentioned, Adversary 2, without access to the original model, inherently leads to lower \nameappend$_{\rm MR}$ attack performance compared to Adversary 1.
Here, we compare with SAMIA~\cite{yuan2022membership} in \nameappend$_{\rm NR}$, as \nameappend$_{\rm NR}$ also avoids using the information from the original model, and SAMIA offers the best performance among \nameappend$_{\rm NR}$.
Table~\ref{composite attack (Adversary 2)} presents the performance improvement of \nameappend$_{\rm MR}$ relative to SAMIA's optimal performance on a single compressed model. Notably, \nameappend$_{\rm MR}$ still maintains significant advantages by leveraging multiple compressed models. 
For example, when all three compression operations are applied, there is a 7.39\% improvement in TPR @ 0.1\% FPR, a 9.7\% increasement in balanced accuracy, and a 14.2\% enhancement in AUC.
This result further emphasizes the superior capability of \nameappend$_{\rm MR}$ in effectively exploiting multiple compressed models to amplify privacy leakage.

\subsection{Discussion}

\mypara{Compression vs Duplication.}
To further illustrate that the superiority of \nameappend$_{\rm SR}$ and \nameappend$_{\rm MR}$ primarily stems from privacy leakage caused by the model compression, rather than simply relying on aggregating leaked information from duplicated models to enhance MIA performance, we conducted a baseline experiment.
Specifically, we use the uncompressed model from previous experiments as the original target model. To simulate a duplicated multi-model setting, we instantiate four different versions of the model with the same architecture but different training hyperparameter settings (e.g., learning rate, training epochs, and random seed). Each of the different settings creates a duplicated model with almost the same testing accuracy as the original model.

As shown in Table~\ref{baseline}, although \nameappend$_{\rm SR}$ outperforms \nameappend$_{\rm NR}$, there remains a significant gap compared to its performance on compression setting. For example, Table~\ref{qat_result} exhibits that \nameappend$_{\rm SR}$ achieves up to 91.1\% attack accuracy on a quantized model, whereas its best performance under the baseline reaches only 72.7\%. 
This is because, while the baseline introduces output variations between the original target model and reference models due to training hyperparameter difference, such differences are relatively small and lack consistent patterns.  
In contrast, model compression substantially alters the model’s capacity and representational behavior, leading to more pronounced and structured differences in outputs, thereby providing stronger discriminative signals for MIA. 
Furthermore, we construct \nameappend$_{\rm MR}$ using four different reference models and observe only marginal improvement over \nameappend$_{\rm SR}$. Thus, these results indicate that the advantage of \name primarily stems from compression-induced privacy leakage, rather than the accumulation of minor variations across multiple duplicated models.

\begin{table}[t]
\centering
\caption{Attack performance of \nameappend$_{\rm MR}$ with different numbers of pruned models on the Mini-ImageNet.}
\label{number of compression models}
\scalebox{0.65}{
    \renewcommand{\arraystretch}{1.2}
    \begin{tabular}{c|ccc}
    \hline
    Number & TPR @ 0.1\% FPR (\%) & Balanced Accuracy (\%)& AUC (\%) \\ 
    \hline
    $L = \{0.6\}$ & 40.2 & 85.8 & 93.9 \\
    $L = \{0.6, 0.7\}$ &44.8 &92.9 &98.0 \\
    $L = \{0.6, 0.7, 0.8\}$ &54.0 &93.6&98.3 \\ 
    $L = \{0.6, 0.7, 0.8, 0.9\}$ &61.3 &94.8 &98.9  \\ 
    
    \hline
    \end{tabular}
}
\end{table}

\subsection{Ablation Study}

\mypara{Impact of the Number of Compressed Models.}
We conduct \nameappend$_{\rm MR}$ (Adversary 1) utilizing 1, 2, 3, and 4 pruned models, respectively, to systematically assess the influence of the number of pruned models on overall attack performance.
As shown in Table~\ref{number of compression models}, the attack performance improves with more pruned models, aligning with the expectation that additional pruned models provide more exploitable leakage information.
For example, using two pruned models increases balanced accuracy by 7.1\% compared to one pruned model.

Given space limitations, the detailed performance of individual components is presented in Appendix~\ref{individual}.

\section{Conclusion}
\label{sec:con}

This work presents the first in-depth privacy evaluation framework \name for three widely used and commercially supported compression operations---pruning, quantization, and weight clustering---through the lens of membership inference.
Specifically, \nameappend$_{\rm SR}$ reveals that these compression operations indeed increase privacy leakage. Notably, this leakage is further exacerbated when leveraging information from multiple compressed models.  Building on this, we propose \nameappend$_{\rm MR}$ that stacks the loss of multiple compressed models and the meta-posterior from the \nameappend$_{\rm SR}$ attack meta-classifier. Extensive experiments have validated the \name superior performance under diverse model architectures, datasets, and various settings.




%
{\footnotesize \bibliographystyle{IEEEtran}
\bibliography{normal_generated_py3}}

\appendix

\subsection{Datasets Description}
\label{appendix:datasets}

Below is a brief description of each dataset.

\mypara{CIFAR-10.}
CIFAR-10~\cite{CIFAR} is a widely used benchmark dataset in image classification, consisting of 60,000 32$\times$32 color images across 10 distinct classes: airplane, automobile, bird, cat, deer, dog, frog, horse, ship, and truck. 
It contains 50,000 training and 10,000 test images, with an equal number of images in each class.

\mypara{CIFAR-100.}
CIFAR-100~\cite{CIFAR} is similar to CIFAR-10 but includes 100 classes instead of 10, with each class containing 600 images. It consists of 60,000 32$\times$32 color images, divided into 50,000 training and 10,000 test images, offering a more granular challenge for image classification models.

\mypara{Mini-ImageNet.} 
Mini-ImageNet~\cite{vinyals2016matching} is a widely used benchmark dataset for evaluating image recognition algorithms. A subset of the ImageNet, it consists of 100 classes, each containing 600 84$\times$84 color images.

\mypara{Tiny-ImageNet.}
Tiny-ImageNet~\cite{le2015tiny} is a widely used benchmark dataset for evaluating image recognition algorithms. It is a subset of the ImageNet, consisting of 100,000 images of 200 categories (500 for each category) downsized to 64×64 colored images. Each category includes 500 training images, 50 validation images.

\mypara{Location.}~\footnote{\url{https://sites.google.com/site/yangdingqi/home/foursquare-dataset}}
This dataset consists of 5,010 samples with 446 binary features and is frequently used in membership inference attacks. The task is to predict a user's geosocial type based on their behavioral records in a 30-class classification problem.

\mypara{Texas.}~\footnote{\url{https://www.dshs.texas.gov/THCIC/Hospitals/Download.shtm}}
This dataset sourced from the Texas Department of State Health Services, includes 67,330 samples and 6,170 binary features related to injuries, diagnoses, procedures, and demographics. The task is to predict one of the 100 most common procedures based on the patient's data.

\subsection{Other Meta-data Construction Methods and Performance}
\label{appendix:sc}

We present two additional meta-data construction methods, similar to~\cite{chen2021machine}. One method is based on direct concatenation, i.e., $\mathcal{P}_{o}^{s} \parallel \mathcal{P}_{c}^{s} \parallel y$,  and the other is based on calculating the $L_{2}$ distance, i.e., $\left\|\mathcal{P}_{o}^{s} - \mathcal{P}_{c}^{s}\right\|_2 \parallel y$. Table~\ref{dirct} provides the attack results for these two construction methods on the clustered MobileNetV2 with 8 clusters on Tiny-ImageNet.

\begin{table}[H]
\centering
\caption{Attack performance of two construction methods on the clustered MobileNetV2 with 8 clusters on Tiny-ImageNet.}
\label{dirct}
\scalebox{0.6}{
\begin{tabular}{c|cccccc}
\toprule
Attack  & \multicolumn{2}{c}{TPR @ 0.1\% FPR (\%)} & \multicolumn{2}{c}{Balanced Accuracy (\%)}  & \multicolumn{2}{c}{AUC (\%)} \\
\cmidrule(r){2-3} \cmidrule(r){4-5} \cmidrule(r){6-7}Method & original &clustered &original &clustered & original &clustered \\     
\toprule
    \nameappend$_{\rm NR}$~\cite{shokri2017membership} (LR)  
    &0.1 & 0.2
    &51.3 &51.7 
    &51.3 &52.3  \\
    \nameappend$_{\rm NR}$~\cite{nasr2018machine} (LR)
    &0.0 &0.0
    &51.5 &51.9 
    &51.6 &52.2   \\
    \nameappend$_{\rm NR}$~\cite{shokri2017membership} (RF)   
    &\textbf{5.0} &5.8
    &62.5 &61.3
    &67.8 &66.2 \\
    \nameappend$_{\rm NR}$~\cite{nasr2018machine} (RF)
    &4.7 &5.5
    &62.2 &61.1 
    &67.7 &66.0  \\
    \nameappend$_{\rm NR}$~\cite{yeom2018privacy}
    &0.0 &0.1
    &55.3 &49.4
    &53.9 &45.6  \\
    \nameappend$_{\rm NR}$~\cite{song2021systematic}
    &0.0 &0.0
    &61.5 &57.1
    &63.4 &57.8  \\
    \nameappend$_{\rm NR}$~\cite{yuan2022membership}   
    &4.9 &5.2
    &\textbf{67.2} &61.0 
    &\textbf{72.6} &67.1  \\
    \midrule
    Direct concatenation (LR)
    &- &0.1
    &- &51.9
    &- &52.4  \\
    $L_{2}$ distance (LR)   
    &- &21.5
    &- &67.8 
    &- &69.6  \\
    Direct concatenation (RF)
    &- &18.1
    &- &\textbf{80.5} 
    &- &\textbf{89.5}  \\
    $L_{2}$ distance (RF) 
    &- &\textbf{36.7}
    &- &69.5
    &- &76.1  \\
\bottomrule
\end{tabular}}
\end{table}

\subsection{The Architecture of FCN}
\label{appendix:fcn}

The FCN's structure as described in the Table~\ref{fcn_architecture}.

\begin{table}[ht]
\centering
\caption{Architecture of the FCN.}
\label{fcn_architecture}

\begin{tabular}{|l|l|l|l|}
\hline
\textbf{Layer} & \textbf{Units} & \textbf{Activation} &\textbf{Regularization} \\ \hline
Layer 1 &(Input, 256) & ReLU & Dropout (0.1) \\ \hline
Layer 2 & (256, 128) & ReLU & Dropout (0.1) \\ \hline
Layer 3& (128, Output) & - & -  \\ \hline
\end{tabular}
\end{table}

\subsection{Evaluation of \nameappend$_{\rm NR}$ and \nameappend$_{\rm SR}$ for other datasets}
\label{appendix:evaluation}

\subsubsection{Evaluation of Pruning}
The attack results for pruning at different sparsities are presented for Location in Table~\ref{location_prune}, Tiny-ImageNet in Table~\ref{tiny_prune}, and CIFAR-10 in Table~\ref{CIFAR-10_prune}.

\begin{table}[H]
\centering
\caption{
Attack performance of different attacks on varying pruned rate (FCN+Location).}
\label{location_prune}
\scalebox{0.6}{
\begin{tabular}{c|ccccccccc}
\toprule
Attack  & \multicolumn{3}{c}{TPR @ 0.1\% FPR (\%)} & \multicolumn{3}{c}{Balanced Accuracy (\%)}  & \multicolumn{3}{c}{AUC (\%)} \\
\cmidrule(r){2-4} \cmidrule(r){5-7} \cmidrule(r){8-10}Method &original & 60\% & 70\% &original & 60\% & 70\% &original & 60\% & 70\% \\ 
    
\toprule
    \nameappend$_{\rm NR}$~\cite{shokri2017membership} (LR) 
    &0.0 &2.9 &2.0
    &49.2 &51.6 &50.2
    &50.7 &52.4 &50.5 \\
    \nameappend$_{\rm NR}$~\cite{nasr2018machine} (LR)
    &0.0 &0.0 &0.0
    &51.7 &54.3 &54.9
    &53.7 &53.5 &52.7 \\
    \nameappend$_{\rm NR}$~\cite{shokri2017membership} (RF) 
    &\textbf{5.9} &5.5  &2.7
    &80.1 &84.3 &81.2
    &91.7 &93.3 &87.9\\
    \nameappend$_{\rm NR}$~\cite{nasr2018machine} (RF)
    &3.5 &4.2 &3.1
    &79.3 &83.8 &80.7
    &91.7 &92.9 &87.2 \\
    \nameappend$_{\rm NR}$~\cite{yeom2018privacy}
    &0.4 &0.1 &0.2
    &81.7 &84.9 &84.7
    &89.5 &90.8 &88.1 \\
    \nameappend$_{\rm NR}$~\cite{song2021systematic}
    &0.3 &0.2 &0.2 
    &\textbf{84.6}&85.0 &87.6  
    &\textbf{90.6} &91.8&89.8  \\
    \midrule
    \nameappend$_{\rm SR}$ 1 (LR)
    &- &0.2 &0.3
    &- &85.1 &87.6 
    &- &91.2 &90.1 \\ 
    \nameappend$_{\rm SR}$ 2 (LR)   
    &- &0.2 &0.1 
    &- &86.3 &88.2
    &- &91.6 &90.4 \\ 
    \nameappend$_{\rm SR}$ 1 (RF)
    &- &6.9 &\textbf{9.3} 
    &- &87.1 &88.2 
    &- &94.1 &93.2\\ 
    \nameappend$_{\rm SR}$ 2 (RF) 
    &- &\textbf{7.2} &9.0
    &- &\textbf{88.0} &\textbf{88.9}
    &- &\textbf{94.4} &\textbf{93.7}\\ 
\bottomrule
\end{tabular}

}
\end{table}

\begin{table*}[!t]
\centering
\caption{
Attack performance of different attacks on varying pruned rate (MobilNetV2+Tiny-ImageNet). }
\label{tiny_prune}
\scalebox{0.7}{
\begin{tabular}{c|ccccccccccccccc}
\toprule
Attack  & \multicolumn{5}{c}{TPR @ 0.1\% FPR (\%)} & \multicolumn{5}{c}{Balanced Accuracy (\%)}  & \multicolumn{5}{c}{AUC (\%)} \\
\cmidrule(r){2-6} \cmidrule(r){7-11} \cmidrule(r){12-16}Method &original & 40\% & 50\% & 60\% & 70\%  &original & 40\% & 50\% & 60\% & 70\% &original & 40\% & 50\% & 60\% & 70\% \\ 
\toprule
    \nameappend$_{\rm NR}$~\cite{shokri2017membership} (LR)     
    &0.2 &0.2 &0.3 &0.1 &0.2
    &51.3 &50.8 &51.0 &51.1 &50.4 
    &51.3 &51.6 &51.0&51.5 &50.9   \\
    \nameappend$_{\rm NR}$~\cite{nasr2018machine} (LR)
    &0.0 &0.0 &0.1 &0.0 &0.1
    &51.5 &51.1 &51.3 &51.0 &50.6
    &51.6 &51.8 &51.9 &51.7 &51.4    \\
    
    \nameappend$_{\rm NR}$~\cite{shokri2017membership} (RF)     
    &\textbf{5.1} &5.3 &5.1 &3.9 &5.2
    &62.5 &61.1 &61.9 &60.9 &60.0
    &67.8 &66.2 &67.3 &65.6 &64.6 \\

    \nameappend$_{\rm NR}$~\cite{nasr2018machine} (RF)
    &4.8 &5.4 &5.0 &4.1 &5.1
    &62.2 &60.9 &61.7 &60.5 &60.1
    &67.7 &65.9 &67.1 &65.3 &64.4 \\

    \nameappend$_{\rm NR}$~\cite{yeom2018privacy}
    &0.1 &0.1 &0.1 &0.0 &0.0
    &55.3 &53.5 &55.6 &52.3 &49.9
    &53.9 &52.0 &54.5 &50.9 &48.0 \\
    
    \nameappend$_{\rm NR}$~\cite{song2021systematic}
    &0.1 &0.1 &0.1 &0.1 &0.1
    &61.5 &60.3 &62.2 &59.9 &57.7
    &63.4 &61.2 &64.4 &61.5 &58.8 \\
    
    \nameappend$_{\rm NR}$~\cite{yuan2022membership}
    &4.9 &3.9 &1.9 &0.1 &0.2
    &\textbf{67.2} &66.3 &67.1 &51.1 &50.1
    &\textbf{72.6} &72.4 &71.8 &55.1 &49.3 \\
    
    \midrule
    \nameappend$_{\rm SR}$ 1 (LR)
    &- &17.9 &14.4 &11.2 &9.8
    &- &68.0 &67.0 &67.4 &67.8
    &- &76.6 &74.6 &75.6 &76.2 \\
    
    \nameappend$_{\rm SR}$ 2 (LR)
    &- &14.5 &11.5 &11.8 &10.5
    &- &68.7 &68.1 &68.8 &69.5
    &- &80.2 &78.7 &79.7 &80.4 \\
    
    \nameappend$_{\rm SR}$ 1 (RF)
    &- &\textbf{54.4} &\textbf{33.8} &\textbf{29.5} &\textbf{24.2}
    &- &91.9 &89.0 &85.2 &81.8
    &- &97.7 &95.9 &92.9 &90.1 \\
    
    \nameappend$_{\rm SR}$ 2 (RF)
    &- &54.0 &\textbf{33.8} &28.8 &\textbf{24.2}
    &- &\textbf{92.1} &\textbf{89.3} &\textbf{85.7} &\textbf{82.7}
    &- &\textbf{97.9} &\textbf{96.2} &\textbf{93.6} &\textbf{91.1} \\

\bottomrule
\end{tabular}}
\end{table*}

\begin{table*}[]
\centering
\caption{Attack performance of different attacks on varying pruned rate (ResNet18+CIFAR-10).}
\label{CIFAR-10_prune}
\scalebox{0.7}{
\begin{tabular}{c|ccccccccccccccc}
\toprule
Attack  & \multicolumn{5}{c}{TPR @ 0.1\% FPR (\%)} & \multicolumn{5}{c}{Balanced Accuracy (\%)}  & \multicolumn{5}{c}{AUC (\%)} \\
\cmidrule(r){2-6} \cmidrule(r){7-11} \cmidrule(r){12-16}Method &original & 60\% & 70\% & 80\% & 90\% &original & 60\% & 70\% & 80\% & 90\% &original & 60\% & 70\% & 80\% & 90\% \\ 
    
\toprule
\nameappend$_{\rm NR}$~\cite{shokri2017membership} (LR)
&0.4 &0.2 &0.2 &0.2 &0.2
&53.0 &49.5 &47.9 &48.1 &49.3
&53.7 &50.0 &48.6 &46.9 &49.9 \\

\nameappend$_{\rm NR}$~\cite{nasr2018machine} (LR)
&0.0 &0.0 &0.0 &0.0 &0.0
&55.2 &49.5 &47.5 &48.0 &46.9
&55.4 &50.0 &50.3 &47.5 &51.5 \\

\nameappend$_{\rm NR}$~\cite{shokri2017membership} (RF)
&0.8 &0.9 &0.6 &0.5 &0.3
&65.6 &68.6 &68.9 &68.5 &61.9
&71.4 &74.3 &74.8 &74.5 &68.2 \\

\nameappend$_{\rm NR}$~\cite{nasr2018machine} (RF)
&\textbf{1.2} &\textbf{1.1} &\textbf{0.7} &\textbf{0.9} &0.3
&68.7 &70.9 &71.4 &71.3 &64.9
&74.6 &76.4 &77.1 &77.1 &71.6 \\

\nameappend$_{\rm NR}$~\cite{yeom2018privacy}
&0.1 &0.2 &0.2 &0.1 &0.1
&65.5 &68.3 &68.6 &68.8 &61.7
&69.1 &71.7 &71.9 &71.9 &67.9 \\

\nameappend$_{\rm NR}$~\cite{song2021systematic}
&0.2 &0.2 &0.2 &0.2 &0.1
&68.4 &70.5 &71.0 &71.2 &68.3
&72.2 &74.0 &74.3 &74.3 &71.5 \\

\nameappend$_{\rm NR}$~\cite{yuan2022membership}
&0.7 &0.4 &0.2 &0.4 &0.2
&\textbf{68.9} &70.9 &71.5 &71.8 &70.2
&\textbf{74.7} &76.3 &76.4 &76.5 &74.9 \\ 
\midrule

\nameappend$_{\rm SR}$ 1 (LR)
&- &0.1 &0.1 &0.1 &0.1
&- &68.2 &67.5 &67.8 &69.0
&- &72.0 &72.1 &72.1 &70.9 \\

\nameappend$_{\rm SR}$ 2 (LR)
&- &0.1 &0.2 &0.1 &0.1
&- &70.3 &70.1 &70.5 &71.2
&- &73.2 &73.5 &73.4 &72.4 \\

\nameappend$_{\rm SR}$ 1 (RF)
&- &0.6 &0.4 &0.5 &0.3
&- &71.1 &70.6 &70.2 &70.1
&- &77.2 &75.3 &75.6 &74.8 \\

\nameappend$_{\rm SR}$ 2 (RF)
&- &0.6 &0.5 &0.5 &\textbf{0.3}
&- &\textbf{72.6} &\textbf{72.5} &\textbf{72.2} &\textbf{72.1}
&- &\textbf{78.4} &\textbf{77.1} &\textbf{77.2} &\textbf{76.8} \\

\bottomrule
\end{tabular}}
\end{table*}

\subsubsection{Evaluation of Quantization}
\label{appendix:quantization}

Here, we provide attack results for other datasets see Table~\ref{quantization}.

\begin{table*}[]
\centering
\caption{Attack performance of quantization.}
\label{quantization}
\scalebox{0.7}{
\begin{tabular}{c|ccccccccc}
\toprule
Attack  & \multicolumn{3}{c}{TPR @ 0.1\% FPR (\%)} & \multicolumn{3}{c}{Balanced Accuracy (\%)}  & \multicolumn{3}{c}{AUC (\%)} \\
\cmidrule(r){2-4} \cmidrule(r){5-7} \cmidrule(r){8-10}Method &Location  &CIFAR-100 &Tiny-ImageNet &Location &CIFAR-100 &Tiny-ImageNet &Location &CIFAR-100 &Tiny-ImageNet\\ 
&FCN &RseNet50 &MobileNetV2 &FCN &RseNet50 &MobileNetV2 &FCN &RseNet50 &MobileNetV2  \\
\toprule
\nameappend$_{\rm NR}$~\cite{shokri2017membership} (LR)     
    &0.2 &0.3 &0.1
    &49.6 &49.7 &51.4
    &50.7 &51.4 &51.6   \\
\nameappend$_{\rm NR}$~\cite{nasr2018machine} (LR)
    &0.0 &0.0 &0.1
    &51.8 &53.4 &51.6
    &53.7 &54.4 &52.3   \\
    
\nameappend$_{\rm NR}$~\cite{shokri2017membership} (RF)     
    &4.1 &1.9 &3.2
    &80.2 &73.4 &61.3
    &91.6 &81.0 &66.5  \\
\nameappend$_{\rm NR}$~\cite{nasr2018machine} (RF)
    &\textbf{5.3} &\textbf{2.2} &3.2
    &79.2 &73.3 &60.9
    &91.5 &80.9 &66.3 \\
\midrule
\nameappend$_{\rm SR}$ 1 (LR) 
    &0.4 &0.1 &28.3
    &83.4 &77.3 &68.6  
    &90.1 &78.8 &77.0    \\
\nameappend$_{\rm SR}$ 2 (LR) 
    &0.4 &0.1 &15.5
    &85.0 &78.1 &67.7 
    &90.7 &79.4 &80.3    \\
\nameappend$_{\rm SR}$ 1 (RF)
    &2.3 &0.5 &60.6
    &84.7 &77.6 &96.7  
    &92.3 &82.6 &99.4    \\
\nameappend$_{\rm SR}$ 2 (RF) 
    &1.4 &0.8 &\textbf{62.9}
    &\textbf{86.3} &\textbf{78.4} &\textbf{97.0}  
    &\textbf{92.8} &\textbf{83.2} &\textbf{99.5}    \\

\bottomrule
\end{tabular}}
\end{table*}

\subsubsection{Evaluation of Weight Clustering}
\label{appendix:cluster}

We employ TF-Lite to perform weight clustering on the original ResNet18 model trained on CIFAR-10, with cluster numbers set to 14 and 8. Note that the original ResNet18 exhibits relatively low overfitting, achieving a train accuracy of 96\% and test accuracy of 85\%, which poses a challenge for MIA. As shown in Table~\ref{CIFAR-10_cluster}, when the meta-classifier structure is LR, \nameappend$_{\rm SR}$ significantly enhances attack performance, surpassing \nameappend$_{\rm NR}$ on either the original or clustered models. Specifically, \nameappend$_{\rm SR}$ improves the balanced accuracy by 6.5\% over \nameappend$_{\rm NR}$~\cite{nasr2018machine} when the cluster number is 8. Similarly, when the meta-classifier structure is RF, \nameappend$_{\rm SR}$ still outperforms \nameappend$_{\rm NR}$ across all model variants. Additionally, we evaluate weight clustering on MobileNetV2 trained on Tiny-ImageNet (implement on PyTorch) with cluster sizes of 8, 16, and 32. The corresponding MIA results are reported in Table~\ref{tiny_cluster}.

\begin{table}[H]
\centering
\caption{Attack performance of different attacks against ResNet18 trained and clustered at various levels on CIFAR-10.}
\label{CIFAR-10_cluster}
\scalebox{0.65}{
\begin{tabular}{c|cccccccccccc}
\toprule
Attack & \multicolumn{3}{c}{Balanced Accuracy (\%)}  & \multicolumn{3}{c}{AUC (\%)} \\
\cmidrule(r){2-4} \cmidrule(r){5-7} Method  &original & 14  & 8 &original & 14  & 8\\ 
\toprule

    \nameappend$_{\rm NR}$~\cite{shokri2017membership} (LR)    
    &49.7 &49.9 &49.8
    &49.2 &50.0 &50.0   \\
    \nameappend$_{\rm NR}$~\cite{nasr2018machine} (LR)
    &49.8 &49.6 &50.1
    &49.5 &48.9 &49.5   \\

    \nameappend$_{\rm NR}$~\cite{shokri2017membership} (RF)    
    &54.1 &54.7 &53.8
    &55.9 &56.8 &55.7   \\
    \nameappend$_{\rm NR}$~\cite{nasr2018machine} (RF)
    &56.0 &56.3 &55.8
    &58.2 &59.0 &58.3   \\
    
    \midrule
    \nameappend$_{\rm SR}$ 1 (LR)
    &- &54.9 &54.8
    &- &54.4 &54.1  \\ 
    \nameappend$_{\rm SR}$ 2 (LR) 
    &- &56.8 &56.6
    &- &56.8 &55.5  \\ 
    
    \nameappend$_{\rm SR}$ 1 (RF)
    &- &54.9 &54.2
    &- &57.2 &56.5  \\ 
    \nameappend$_{\rm SR}$ 2 (RF) 
    &- &\textbf{57.0} &\textbf{56.8} 
    &- &\textbf{59.5} &\textbf{59.6}  \\ 
\bottomrule
\end{tabular}}
\end{table}

\begin{table*}[]
\centering
\caption{Attack performance of different attacks against MobileNetV2 trained and clustered at various levels on Tiny-ImageNet.}
\label{tiny_cluster}
\scalebox{0.7}{
\begin{tabular}{c|cccccccccccc}
\toprule
Attack  & \multicolumn{4}{c}{TPR @ 0.1\% FPR (\%)} & \multicolumn{4}{c}{Balanced Accuracy (\%)}  & \multicolumn{4}{c}{AUC (\%)} \\
\cmidrule(r){2-5} \cmidrule(r){6-9} \cmidrule(r){10-13}Method &original & 32 & 16 & 8 &original & 32 & 16 & 8 &original & 32 & 16 & 8 \\ 
\toprule
\nameappend$_{\rm NR}$~\cite{shokri2017membership} (LR)     
    &0.2 &0.1 &0.3 &0.2
    &51.3 &50.9 &51.9 &51.7
    &51.3 &52.0 &52.3 &52.3   \\
\nameappend$_{\rm NR}$~\cite{nasr2018machine} (LR)
    &0.0 &0.0 &0.0 &0.1
    &51.5 &51.5 &51.5 &51.9
    &51.6 &52.0 &52.1 &52.2   \\
    
\nameappend$_{\rm NR}$~\cite{shokri2017membership} (RF)     
    &\textbf{5.1} &5.3 &6.3 &5.8
    &62.5 &62.2 &62.3 &61.3
    &67.8 &67.3 &67.7 &66.2 \\
\nameappend$_{\rm NR}$~\cite{nasr2018machine} (RF)
    &4.8 &5.2 &6.3 &5.5
    &62.2 &61.9 &62.1 &61.1
    &67.7 &67.0 &67.5 &66.0\\

\nameappend$_{\rm NR}$~\cite{yeom2018privacy}
    &0.1 &0.1 &0.1 &0.1
    &55.3 &55.7 &53.6 &49.4
    &53.9 &54.1 &51.8 &45.6 \\
\nameappend$_{\rm NR}$~\cite{song2021systematic}
    &0.1 &0.1 &0.1 &0.1
    &61.5 &61.7 &59.6 &57.1
    &63.4 &63.9 &62.2 &57.8  \\
\nameappend$_{\rm NR}$~\cite{yuan2022membership}   
    &4.9 &4.2 &4.9 &5.2 
    &\textbf{67.2} &67.0 &66.0 &61.0 
    &\textbf{72.6} &73.0 &72.5 &67.1  \\ 
    
\midrule
\nameappend$_{\rm SR}$ 1 (LR) 
    &- &27.7 &22.4 &17.0
    &- &69.0 &69.2 &68.2 
    &- &77.3 &77.4 &76.5    \\
\nameappend$_{\rm SR}$ 2 (LR) 
    &- &14.9 &11.8 &12.0
    &- &68.0 &69.2 &69.6  
    &- &80.4 &80.9 &80.5   \\
\nameappend$_{\rm SR}$ 1 (RF)
    &- &87.2 &55.5 &28.7
    &- &\textbf{96.9} &\textbf{93.0} &85.9  
    &- &99.6 &98.2 &93.7    \\
\nameappend$_{\rm SR}$ 2 (RF) 
    &- &\textbf{88.2} &\textbf{59.3} &\textbf{30.6}
    &- &96.7 &\textbf{93.0} &\textbf{86.3}  
    &- &\textbf{99.6} &\textbf{98.3} &\textbf{94.3}    \\

\bottomrule
\end{tabular}}
\end{table*}

\subsection{The Impact of the Victim's Dataset}
\label{appendix:victim_dataset}

\mypara{Dataset of the Victim Model.}
Here, we focus on examining the influence of datasets on attacks by using the same model architecture with different datasets. Specifically, we trained the VGG16 model on Mini-ImageNet and CIFAR-10. As shown in Table \ref{victim dataset}, the attack performance of the \nameappend$_{\rm NR}$ on both the original model and the pruned model is consistently lower for Mini-ImageNet compared to CIFAR-10, regardless of the attack meta-model used. 
However, with our \nameappend$_{\rm SR}$, which leverages the variation introduced by the compression operation, the attack performance on Mini-ImageNet is significantly higher than on CIFAR-10. 
We hypothesize that this is because Mini-ImageNet is a more complex dataset than CIFAR-10, causing larger compression-induced variation differences between members and non-members. 
Therefore, under the same model architecture, compressing more complex datasets can lead to more severe privacy leakage.

\begin{table}[H]
\centering
\caption{The attack performance of CIFAR-10/Mini-ImageNet on VGG16.}
\label{victim dataset}
\scalebox{0.58}{
\begin{threeparttable}
\begin{tabular}{c|cccccc}
\toprule
Attack  & \multicolumn{3}{c}{Balanced Accuracy (\%)}  & \multicolumn{3}{c}{AUC (\%)} \\
\cmidrule(r){2-4} \cmidrule(r){5-7} Method &original & pruned (80\%) & quantized &original & pruned (80\%) & quantized \\ 
    
\toprule
\nameappend$_{\rm NR}$~\cite{shokri2017membership} (LR)
    &50.3/48.3 &50.5/48.5 &50.8/48.3
    &50.1/47.3 &50.3/48.0 &50.3/47.4\\
\nameappend$_{\rm NR}$~\cite{nasr2018machine} (LR)
    &49.8/48.3 &50.1/50.0 &50.0/51.1
    &49.6/50.1 &50.2/51.0 &50.0/50.5\\
\nameappend$_{\rm NR}$~\cite{shokri2017membership} (RF)
    &61.7/59.3 &59.7/58.6 &61.6/59.4
    &65.9/63.2 &63.0/62.1 &65.9/63.4\\
\nameappend$_{\rm NR}$~\cite{nasr2018machine} (RF) 
    &61.8/59.2 &60.5/58.7 &61.9/59.3
    &66.4/63.3 &63.9/62.3 &66.2/63.5\\
\midrule
\nameappend$_{\rm SR}$ 2 (LR) 
    &- &62.1/59.9 &61.8/60.8
    &- &61.2/68.8 &61.1/70.8\\
\nameappend$_{\rm SR}$ 2 (RF)
    &- &\textbf{62.4/83.4} &62.1/90.3
    &- &\textbf{66.6/92.1} &\textbf{66.4/98.3}
    \\

\bottomrule
\end{tabular}
\end{threeparttable}
}
\end{table}

\subsection{Attack performance on baseline}
The attack results for the baseline methods are presented in Table~\ref{baseline}.
\begin{table*}[t]
\centering
\caption{
Attack performance on baseline (VGG16+Mini-ImageNet).}
\label{baseline}
\scalebox{0.8}{
\begin{threeparttable}
\begin{tabular}{c|ccccc>{\columncolor{gray!20}}cccccc>{\columncolor{gray!20}}cccccc>{\columncolor{gray!20}}c}
\toprule
Attack  & \multicolumn{6}{c}{TPR @ 0.1\% FPR (\%)} & \multicolumn{6}{c}{Balanced Accuracy (\%)}  & \multicolumn{6}{c}{AUC (\%)} \\
\cmidrule(r){2-7} \cmidrule(r){8-13} \cmidrule(r){14-19}Method &original & R1 & R2 & R3 & R4 &prune 
&original & R1 & R2 & R3 & R4 &prune 
&original & R1 & R2 & R3 & R4 &prune  \\ 
    
\toprule
    \nameappend$_{\rm NR}$~\cite{shokri2017membership} (RF) 
    &1.6 &1.0 &1.1 &1.2 &1.1 &1.5
    &59.3 &58.4 &59.7 &59.1 &59.6 &59.1
    &63.2 &62.5 &63.7 &63.0 &63.8 &62.7\\
    \nameappend$_{\rm NR}$~\cite{nasr2018machine} (RF)
    &1.5 &1.1 &1.0 &1.0 &1.0 &1.4
    &59.2 &58.2 &59.4 &58.7 &59.3 &59.2
    &63.3 &62.3 &63.3 &62.7 &63.4 &62.8\\
    \midrule
    \nameappend$_{\rm SR}$ 1 (RF)
    &- &12.4 &15.0 &15.8 &13.3 &42.1
    &- &72.2 &71.9 &72.0 &71.8 &84.2
    &- &81.0 &80.9 &81.0 &80.9 &92.8 \\
    \nameappend$_{\rm SR}$ 2 (RF) 
    &- &12.6 &16.3 &16.6 &13.6 &41.7
    &- &72.7 &72.6 &72.7 &72.3 &83.9
    &- &81.8 &81.7 &81.9 &81.7 &93.0\\
    \midrule
    \nameappend$_{\rm MR}$ 
    & & &20.6 & & &61.3
    & & &73.1 & & &94.8
    & & &83.0 & & &98.9\\
\bottomrule
\end{tabular}
\begin{tablenotes} 
    \item The training and test accuracies of the four reference models are as follows: R1 (90.58\%, 75.13\%), R2 (92.31\%, 75.39\%), R3 (90.90\%, 75.30\%), and R4 (92.85\%, 75.78\%).
    \item As a comparison, \colorbox{gray!20}{prune} report results under 70\% pruning for \nameappend$_{\rm NR}$ and \nameappend$_{\rm SR}$, and aggregate multiple pruning rates (60\%, 70\%, 80\%, and 90\%) for \nameappend$_{\rm MR}$.
\end{tablenotes} 
\end{threeparttable}
}
\end{table*}

\subsection{Performance of Individual Component}
\label{individual}

As \nameappend$_{\rm MR}$ utilizes the loss concatenation and posterior concatenation to form meta-data.
To validate the contribution of each meta-data component, we evaluate attack performance using each component individually.
As shown in Table~\ref{component}, the contribution of posterior concatenation is the most pronounced, and the combination of both components surpasses the performance of any individual component.
This underscores the essential contribution of each component to the overall attack performance.

\begin{table}[H]
\caption{Attack performance of \nameappend$_{\rm MR}$ using the individual meta-data component under Adversary 1. }
\label{component}
\scalebox{0.8}{
    \begin{tabular}{c|ccc}
    \hline
    Component & TPR @ 0.1\% FPR (\%) & Balanced Accuracy (\%)& AUC (\%)\\ 
    \hline
    Loss &1.0 &72.0 &73.7 \\
    Posterior&55.6  &93.6  &98.3 \\
    Loss \& Posterior &61.3 &94.8 &98.9 \\
    \hline
    \end{tabular}
}
\end{table}

\subsection{Other Results}
\begin{table}[t]
\caption{Classification accuracy of the 80\% pruned ResNet-18 fine-tuned with different ${N}_{f}$ for CIFAR-10.}
\label{acc_f}
\centering
\scalebox{0.8}{
\begin{tabular}{c|ccc}
\toprule

& \multicolumn{3}{c}{${N}_{f}$} \\
\cmidrule(r){2-4} Dataset  & 90\% & 50\% & 10\% \\ 

\midrule
$\mathcal{D}_{f}$            & 99.80 & 99.85 & 99.95 \\
$\mathcal{D}_{nf}$           & 99.65 & 99.05 & 98.35 \\
Test Dataset & 88.50 & 88.70 & 86.45 \\ 
\bottomrule
\end{tabular}}
\end{table}

\end{document}